\documentclass[a4paper,12pt]{article}
\usepackage{latexsym,amsmath,amsfonts,amssymb}
\usepackage[latin1]{inputenc}
\usepackage[american]{babel}
\usepackage[dvips]{graphicx}
\usepackage{bbm}

\newcommand{\Nugual}[1]{$\mathcal{N}= #1 $}

\numberwithin{equation}{section}

\newcommand{\be}{\begin{equation}} \newcommand{\ee}{\end{equation}}
\newcommand{\bea}{\begin{equation} \begin{aligned}} \newcommand{\eea}{\end{aligned} \end{equation}}

\newcommand{\calC}{\mathcal{C}}
\newcommand{\calD}{\mathcal{D}}

\newcommand{\calL}{\mathcal{L}}

\newcommand{\calN}{\mathcal{N}}
\newcommand{\calO}{\mathcal{O}}
\newcommand{\calQ}{\mathcal{Q}}
\newcommand{\calR}{\mathcal{R}}
\newcommand{\calS}{\mathcal{S}}

\newcommand{\bbC}{\mathbb{C}}

\newcommand{\bbR}{\mathbb{R}}
\newcommand{\bbZ}{\mathbb{Z}}

\newcommand{\eps}{\epsilon}

\begin{document}

\makeatletter \@addtoreset{equation}{section} \makeatother
\renewcommand{\theequation}{\thesection.\arabic{equation}}
\pagestyle{empty}

\rightline{TAUP-2896/09}
\vspace{0.8cm}
\begin{center}
{\LARGE{\bf Transmutation of  $\mathbf{{\cal N}=2}$ fractional D3 branes into twisted sector fluxes \\[15mm]}} {\large{Stefano Cremonesi \\[5mm]}}
{\small{{}
Raymond and Beverly Sackler School of Physics and Astronomy \\
 Tel-Aviv University, Ramat-Aviv 69978, Israel\\
\medskip \tt{stefano@post.tau.ac.il}
}}

\bigskip

\bigskip

\bigskip

\bigskip

\bigskip

{\bf Abstract}
\vskip 20pt
\begin{minipage}[h]{16.0cm}

We study the prototype of fractional D3 branes at non-isolated singularities in gauge/gravity duality at the nonperturbative level.
We embed the quantum moduli space of $\calN=2$ pure SYM, the gauge theory on fractional D3 branes at the $A_1$ singularity, into that of the cascading quiver gauge theory on regular and fractional D3 branes at the same singularity, for which a gravity dual description exists. 
We deduce a simple analytic expression for the exact twisted sector fields in the type IIB string dual, which encodes the full quantum dynamics of the gauge theory. 
Nonperturbative effects in the gauge theory translate into the transmutation of fractional D3 branes into twisted sector fluxes.

\end{minipage}
\end{center}
\newpage
\setcounter{page}{1} \pagestyle{plain}
\renewcommand{\thefootnote}{\arabic{footnote}} \setcounter{footnote}{0}

\tableofcontents

\vspace*{1cm}


\section{Introduction}

Fractional D3 branes have played an important r\^ole in extending AdS$_5$/CFT$_4$ dualities to settings where the gauge theory is not scale invariant. Being nothing but D5 branes wrapped on collapsed 2-cycles which exist at Calabi-Yau (CY) threefold conical singularities, they source 3-form fluxes in the geometry, which then lead to a logarithmically varying 5-form flux. The field-theoretic dual interpretation involves a cascading renormalization group (RG) flow, where the number of degrees of freedom decreases at subsequent strong coupling transitions, until the low energy gauge theory on the worldvolume of fractional D3 branes is reached in the deep infrared. 

In the past few years the attention has been mainly drawn to D5 branes wrapped on rigid collapsed 2-cycles, the best known examples of which are fractional D3 branes at the tip of the conifold \cite{Klebanov:2000hb} and of the complex cone over the first Del Pezzo surface \cite{Herzog:2004tr}: there the gauge theories have $\calN=1$ supersymmetry, the cascade is an infinite sequence of Seiberg dualities \cite{Strassler:2005qs}, and the low energy confining dynamics drives either chiral symmetry breaking \cite{Klebanov:2000hb} or a runaway \cite{DSB,Intriligator:2005aw,Argurio:2007vq}.

That focus was motivated by the fact that $\calN=1$ SYM is a quite close but yet controllable relative of theories of phenomenological interests like pure YM theory and QCD, and theories with a runaway were originally hoped to provide a starting point in the search of gravity dual descriptions of supersymmetry breaking vacua \cite{DSB,Argurio:2006ew}.

There is another class of fractional D3 branes, called of $\calN=2$ kind, which are D5 branes wrapped on exceptional 2-cycles living at non-isolated singularities. The holomorphic data of their macroscopic dynamics is analogous to that of $\calN=2$ SYM, with a Coulomb branch of supersymmetric vacua. Despite having been introduced long ago in the context of gauge/string duality \cite{Klebanov:1999rd,Bertolini:2000dk,Polchinski:2000mx,Billo:2001vg} following pioneering works on D-branes on orbifolds \cite{Douglas:1996sw,Diaconescu:1997br,Billo:2000yb} and their embedding in AdS/CFT \cite{Kachru:1998ys}, only very recently the field-theoretic interpretation of the type IIB near-horizon backgrounds generated by the backreaction of $\calN=2$ fractional D3-branes at the $\bbC\times\bbC^2/\bbZ_2$ orbifold singularity was fully teased out \cite{Benini:2008ir}, settling a long-standing issue in the literature \cite{Polchinski:2000mx,Aharony:2000pp,Petrini:2001fk}. The cascade is understood in this case as a sequence of strong coupling transitions reminiscent of the transition between high energy and low energy theory at the baryonic root of $\calN=2$ SQCD \cite{Argyres:1996eh}.

More complicated CY singularities generically contain rigid as well as non-rigid collapsed 2-cycles. The infinite cascade which UV-completes the low energy dynamics on a generic stack of fractional D3 branes at one of such singularities while allowing us to remain in the realm of gauge/gravity duality consists of a sequence of strong coupling transitions, some of which are Seiberg dualities and others of which are $\calN=2$ baryonic root transitions. This was confirmed in a specific example by studying the behavior of Page charges under cascade transitions on the supergravity side of the duality \cite{Argurio:2008mt}.

\begin{figure}[tn]
\centering
\hspace{1cm}
\includegraphics[width=.5\textwidth]{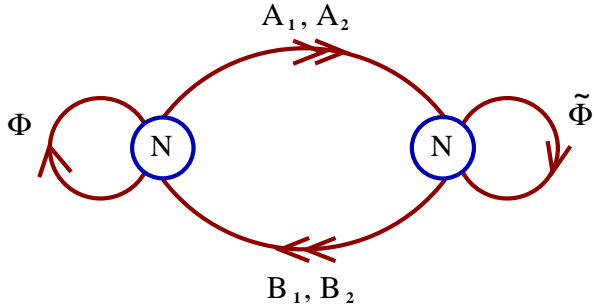}
\caption{\small Quiver diagram of the $U(N) \times U(N)$ \Nugual{2} theory on regular D3 branes at the $\bbC\times\bbC^2/\bbZ_2$ orbifold, in \Nugual{1} notation.
Nodes represent unitary gauge groups, arrows connecting different nodes represent bifundamental chiral superfields, while arrows going from one node to itself represent adjoint chiral superfields. The superpotential is dictated by $\calN=2$ supersymmetry.
\label{fig:N=2_quiver}}
\end{figure}
This paper refines and extends the analysis of the cascading theory on regular and fractional D3 branes at the $\bbC\times\bbC^2/\bbZ_2$ orbifold carried out in \cite{Benini:2008ir} on both sides of the duality: the gauge theory side, whose quiver is depicted in the conformal case in Figure \ref{fig:N=2_quiver}, and the `gravity' side, with a metric, a RR 5-form, and a holomorphic complex scalar that has to be supplemented to account for additional massless modes arising from the twisted sector of closed string theory on the orbifold. The field strength of the complex scalar is the reduction of the complexified 2-form potential on the exceptional 2-cycle of the orbifold.

The aim of the paper is twofold. 
We will first describe in section \ref{sec:embedding} how to embed the moduli space of the gauge theory on fractional D3 branes in the Coulomb branch of the quiver gauge theory with an infinite cascade via their Seiberg-Witten (SW) curves. This embedding will tighten the analogy between the strong coupling transitions in the cascade to the high-low energy transitions at the baryonic root of $\calN=2$ SQCD \cite{Argyres:1996eh}, since the branch points of the curve that are related to the cascade transitions will be exactly double, like the branch points at the baryonic root. It will also allow us to find the exact twisted field configurations in the type IIB duals of those vacua, which encode the full nonperturbative dynamics on the gauge theory side. In section \ref{sec:examples} we will study those twisted field solutions for some interesting vacua in the infinitely cascading theory: we will first consider the $\bbZ_{2M}$-symmetric enhan\c con vacuum originally studied in the literature, and compare it with the approximation used in \cite{Benini:2008ir}; next we will analyze one of the $M$ vacua whose SW curves have genus zero, and which flow to the $M$ vacua of the Klebanov-Strassler theory upon mass deformation.

In the second part of the paper we will employ the previous results to infer properties of fractional D3 branes at nonvanishing string coupling. In section \ref{sec:dissolution} we will show that fractional D3 branes transmute into twisted fields as soon as the string coupling is switched on. Their D5 and D3 brane charges are entirely provided by the twisted fields, with no additional sources.
This is the translation to the string side of the duality of the well-known splitting of classically double branch points in the SW curve, driven by nonperturbative effects in the gauge theory, or equivalently the T-dual manifestation of suspended D4 branes in type IIA string theory becoming M5 brane tubes as soon as the string coupling does not vanish.
We will show how this phenomenon solves two interrelated issues of divergences in the D3 brane charge and the warp factor which appeared in the naive type IIB solutions.
In section \ref{sec:HvsC} we will generalize the observations made in the previous section to generic points of the moduli space of the quiver gauge theory on D3 branes at the  $\bbC\times\bbC^2/\bbZ_2$ orbifold, with or without cascade. We will see that the so-called H-picture of \cite{Benini:2008ir}, where the integral of the NSNS $B_2$ potential on the exceptional 2-cycle remains bounded in the solution and alternates regimes of growth and decrease along the cascade, is generally singled out against the so-called C-picture, where the integral is monotonic along the cascade and diverges at large radii in the limit of infinite cascade. However, the C-picture turns out to be valid (and equivalent to the H-picture) when there is an infinite cascade with exactly double branch points associated to it in the SW curve, as in the vacua studied in the first part of the paper.

We end the paper with a short summary and conclusions. We added an appendix with a discussion of a particular twisted field configuration for fractional branes at an orbifold of the conifold, which turns out to be directly related to one of the configurations studied in the body of the paper.


\section{Embedding the moduli space of pure SYM into the Coulomb branch of a cascading quiver theory}
\label{sec:embedding}

Supergravity%
\footnote{With an abuse of language we will call `supergravity' the low energy theory describing the interactions of all the massless modes of closed string theory, including modes in the twisted sector in the case of orbifolds.}
solutions dual to quiver gauge theory vacua whose RG flows involve an infinite cascade can be found by computing the backreaction of a (large) number of fractional D3 branes at Calabi-Yau conical singularities in the near-horizon limit, with no need of adding the regular D3 branes of the AdS$_5$/CFT$_4$ correspondence. 
One could naively expect that backgrounds found in this way describe holographically the low energy field theory on the fractional D3 branes: for instance $\calN=2$ or $\calN=1$ pure SYM, or more complicated nonconformal gauge theories whose content depends on properties of the singularity and the kind of fractional D3 branes. This point of view was indeed taken in \cite{Bertolini:2000dk}, where a background sourced by $M$ fractional D3 branes at the $A_1$ singularity was originally interpreted as dual to a vacuum of the $\calN=2$ $SU(M)$ SYM theory hosted by the fractional branes. Such low energy theories, however, always contain asymptotically free gauge groups, and therefore by the common lore of gauge/string duality they are not expected to have weakly curved gravity duals, as opposed to what is found. This apparent contradiction is overcome by realizing that the type IIB supergravity solutions, having a finite and constant axio-dilaton, actually describe vacua of the quiver gauge theories living on regular and fractional D3 branes; an infinite cascade completes the gauge theory on fractional branes in the ultraviolet, keeping the gauge couplings bounded from below according to the value of the axio-dilaton.

Moduli spaces of cascading quiver gauge theories are obviously much richer than those of the theories describing their infrared regimes: they can include not only a larger Coulomb branch, parametrized by displacements of fractional D3 branes of all possible kinds, but also a Higgs branch, parametrized by displacements of regular D3 branes.%
\footnote{If the cascade is infinite these branches are infinite-dimensional. When all the fractional D3 branes are of $\calN=2$ kind, the IR dynamics of interest can also arise from a UV conformal quiver gauge theory hosted by regular D3 branes alone, making the moduli space finite-dimensional.}
However, the moduli space of the fractional brane gauge theory can be naturally embedded into the moduli space of the quiver gauge theories of regular and fractional D3 branes with infinite cascade. The background found by backreacting $M$ fractional D3 branes is not dual to a vacuum of the fractional brane theory, but rather to its embedding into the infinitely cascading quiver gauge theory of regular and fractional D3 branes. 

This statement may look trivial for the gauge theories on fractional branes at isolated singularities, for which only a finite number of supersymmetric vacua exists after the complex structure deformation takes place:
 it is well known for instance that one can associate to each of the $M$ vacua of $\calN=1$ $SU(M)$ pure SYM  a vacuum of the cascading Klebanov-Strassler theory and a dual background \cite{Klebanov:2000hb}. 
The content of the previous statement considerably increases when fractional D3 branes at non-isolated singularities are involved: then it becomes a statement about the embedding of the whole $(M-1)$-dimensional moduli space of the fractional brane theory into the infinite-dimensional Coulomb branch of the quiver theory with infinite cascade.

Extending results of \cite{Benini:2008ir}, in this section we will explicitly provide such an embedding for the case of the gauge theory hosted by D3 branes at the $A_1$ singularity (namely the orbifold $\bbC\times\bbC^2/\bbZ_2$) in type IIB string theory. By means of Seiberg-Witten theory, its M theory realization, and the duality of M theory to type IIB string theory, we will also be able to provide the \emph{exact} type IIB twisted fields for that infinite class of vacua in analytic form. The warp factor is then determined from the twisted fields via a 2-dimensional integral on the orbifold fixed plane.

In the following section we will first apply these exact results to some of the vacua studied in \cite{Benini:2008ir}, and then we will exploit them to unravel the fate of fractional D3 branes at nonvanishing string coupling in sections \ref{sec:dissolution} and \ref{sec:HvsC}.

\subsection{Seiberg-Witten curves}

Seiberg-Witten curves manifest themselves in M theory as holomorphic embeddings of M5 branes \cite{Witten:1997sc}, which are the uplifts (with a rescaling) of systems of D4 branes suspended between parallel NS5 branes in type IIA string theory. The low energy gauge dynamics on the suspended D4 branes is a 4-dimensional $\calN=2$ gauge theory,  whose full quantum dynamics is encoded by the M theory uplift.
The $\calN=2$ pure gauge theory is engineered by an M5 brane spanning $\bbR^{1,3}$ and a Riemann surface in $\bbR^2\times$ cylinder, and located at a point in the three additional dimensions. $\calN=2$ supersymmetry requires the embedding to be holomorphic with respect to complex coordinates $v$ and $u$ on $\bbR^2$ and the cylinder respectively. If instead of a cylinder we consider a 2-torus, we are led to the $\calN=2$ quiver theory with two gauge groups coupled by two bifundamental hypermultiplets, like the one of Figure \ref{fig:N=2_quiver}, possibly with different ranks.

The embedding of the whole moduli space of the \Nugual{2} $SU(M)$ pure SYM theory into the moduli space of the quiver gauge theory with an infinite cascade follows from the M theory construction, as we now lay out, generalizing the analysis carried out in \cite{Benini:2008ir} for genus zero SW curves.
We start with \Nugual{2} $SU(M)$ pure SYM theory. Each point of its moduli space is characterized by the
SW curve fibered over it, which in terms of dimensionless variables looks as follows \cite{SU(N),Witten:1997sc}: 
\begin{equation}\label{SW_curve_glue_dimless}
t-P_M(v)+\frac{1}{t}=0\;,
\end{equation}
where 
\begin{equation}
P_M(v)\equiv\prod_{i=1}^M(v-v_i)
\end{equation}
is the characteristic polynomial of the adjoint scalar (in units of the nonperturbative scale $\Lambda$).
The quiver gauge theory is realized in M theory as an elliptic model, defined by the torus identification 
\begin{equation}\label{torus_identification_u}
u \equiv i\frac{x^6+ix^{10}}{2\pi R_{10}}  \sim u+1\sim u+\tau\;,
\end{equation}
or equivalently 
\begin{equation}\label{torus_identification_t}
t \equiv e^{2\pi i u} \sim qt\;,\qquad\qquad q\equiv e^{2\pi i\tau}\;.
\end{equation}

The embedding of the moduli space of \Nugual{2} $SU(M)$ pure SYM theory into the moduli space of the quiver gauge theory with an infinite cascade is easily obtained by wrapping the SW curves \eqref{SW_curve_glue_dimless} infinitely many times on the torus \eqref{torus_identification_u}-\eqref{torus_identification_t}:
\begin{equation}
\begin{split}
0 &= \tilde Q (t,P_M(v))=\lim_{K\to\infty} \tilde Q_K (t,P_M(v))\;,\\
\tilde Q_K (t,P_M(v))&= q^{K(K+1)} f(q)\prod_{j=-K}^K  \left(q^j t - P_M(v) + \frac{1}{q^jt}\right)\;,
\end{split}
\end{equation}
where the $q^{K(K+1)}$ factor is needed for convergence as $K\to\infty$ and 
\begin{equation}
f(q)\equiv\prod_{l=1}^\infty (1-q^{2l})(1-q^{2l-1})^2
\end{equation}
is put for later convenience. The $K\to\infty$ limit converges for any $t$ and $P_M(v)$ because $|q|<1$, and we get the curve%
\footnote{By construction, the locus of solutions of equation \eqref{SW_curve_inf_cascade} for the SW curve wrapped on the torus consists of the two roots in $t$ of equation \eqref{SW_curve_glue_dimless} for the SW curve defined on the cylinder, along with all their infinitely many images under the $t\sim q t$ equivalence which defines the M theory torus as a quotient of the cylinder.} 
\begin{equation}\label{SW_curve_inf_cascade}
\begin{split}
\tilde Q (t,P_M(v)) &= f(q)\,\left(t-P_M(v)+\frac{1}{t}\right) \cdot\\
 &\cdot \prod_{j=1}^\infty \left(
1 - P_M(v) t q^j + t^2 q^{2j} \right) \left( 1 - \frac{P_M(v)}{t} q^j +
\frac{q^{2j}}{t^2} \right) = 0\;.
\end{split}
\end{equation}
To the aim of proving that \eqref{SW_curve_inf_cascade} is a legitimate SW curve for the quiver gauge theory with an infinite cascade in the UV, we then define a sequence (in $K$) of SW curves for the $SU((2K+1)M)\times SU((2K+1)M)$ quiver theory with equal bare complexified gauge couplings \cite{Ennes:1999fb,Petrini:2001fk}
\begin{equation}\label{SW_quiver_K}
\mathcal{Q}_K(t,P_M(v))\equiv q^{K(K+1)}\left[-\calR_K(v)\theta_3(2u|2\tau)+\mathcal{S}_K(v)\theta_2(2u|2\tau)\right]=0\;,
\end{equation}
with suitable characteristic polynomials of the adjoint scalars:
\begin{equation}\label{characteristic_polynomials}
\begin{split}
\calR_K(v) &= P_M(v) \prod_{j=1}^K \left[
P_M(v)^2 +\frac{(1-q^{2j})^2}{q^{2j}} \right] \\
\calS_K(v) &= \left(P_M(v)+q^{-K-\frac{1}{4}}\right) \prod_{j=1}^K \left[
P_M(v)^2 +\frac{(1-q^{2j-1})^2}{q^{2j-1}} \right]
\end{split}
\end{equation}
The $K\to\infty$ limit converges for any $t=e^{2\pi i u}$ and $P_M(v)$ since $|q|<1$, and we call 
\begin{equation} \label{SW_curve_inf_cascade_quiver}
\calQ(t,P_M(v))=\lim_{K\to\infty }\calQ_K(t,P_M(v))\;.
\end{equation}
It is then possible to verify, as was done in \cite{Benini:2008ir}, that for any choice of $t$ and $P_M(v)$
\begin{equation}
\tilde Q(t,P_M(v))= \calQ(t,P_M(v)) \;. 
\end{equation}
Therefore the holomorphic curve \eqref{SW_curve_inf_cascade} obtained by wrapping a SW curve of the pure SYM theory arises as the infinite cascade limit \eqref{SW_curve_inf_cascade_quiver} of a sequence of specific SW curves \eqref{SW_quiver_K}-\eqref{characteristic_polynomials} for the quiver theory on regular D3 branes at the $\bbC\times\bbC^2/\bbZ_2$ orbifold.

This construction holds for any choice of the characteristic polynomial $P_M(v)$; hence it provides the promised embedding of the whole moduli space of $\calN=2$ $SU(M)$ SYM into the infinite-dimensional Coulomb branch of the cascading quiver gauge theory on D3 branes at the $\bbC\times\bbC^2/\bbZ_2$ orbifold singularity.


\subsection{Exact twisted field configurations in type IIB string theory}

The type IIB supergravity solutions dual to the infinite class of cascading vacua previously discussed are easily obtained, by recalling that M theory compactified on a torus of complex structure $\tau$ is equivalent, in the zero size limit of the torus, to type IIB string theory with axio-dilaton $C_0+\frac{i}{g_s}=\tau$.
The main character in the type IIB solutions is the twisted sector complex scalar
\begin{equation} \label{gamma_def}
\gamma\equiv c+\tau b = 
\frac{1}{4\pi^2\alpha'}\int_\calC \left( C_2 + \tau B_2 \right) \;.
\end{equation}
$\calC$ is the exceptional 2-cycle living at the orbifold fixed plane, and $C_2$ and $B_2$ RR and NSNS potentials. Supersymmetry requires that $\gamma$ be a holomorphic function of the complex length coordinate $z$ on the orbifold fixed plane, which is related to $v$ in M theory as $z = 2\pi \alpha'\Lambda\,v$. We will use the dimensionless $v$ instead of $z$ in the remainder of the article.
By duality, the twisted sector complex scalar in type IIB is given by the distance vector between the two branches of the M5 brane on the torus:
\begin{equation}\label{gamma_from_u}
\gamma(v)= u_-(v)-u_+(v)\;,
\end{equation}
where 
\begin{equation}\label{u_from_t}
u_\pm (v) \equiv \frac{1}{2\pi i}\log t_\pm(v)
\end{equation}
and $t_\pm(v)$ are the two solutions of \eqref{SW_curve_glue_dimless} at fixed $v$:%
\footnote{We used the unwrapped curve \eqref{SW_curve_glue_dimless} instead of the wrapped one \eqref{SW_curve_inf_cascade} solely as a simplifying choice, since the information they encode is the same. Picking any other pair of solutions of \eqref{SW_curve_inf_cascade} and \eqref{u_from_t} which are not equivalent under \eqref{torus_identification_t} leads to the same result for the twisted sector scalar \eqref{gamma_from_u} up to its periodicities $\gamma\sim\gamma+1\sim\gamma+\tau$ which amount to large gauge transformations.}
\begin{equation}\label{t_pm}
t_\pm (v)= \frac{1}{2} \left[ P_M(v) \pm \sqrt{P_M(v)^2-4} \right]\;.
\end{equation}
Note that we implicitly chose the C-picture of \cite{Benini:2008ir}, which is naturally suited for infinite cascades with exactly double branch points, as those under discussion. In section \ref{sec:HvsC} we will discuss the H-picture, which turns out to be valid more generally, and motivate why the C-picture can be used to describe this class of vacua as well.

The previous result can be recast in the form
\begin{equation}\label{gamma_generic_poly}
\gamma(v)=\frac{i}{\pi} \log \left[ \frac{P_M(v)}{2}+ \sqrt{\left(\frac{P_M(v)}{2}\right)^2 -1} \right]= \frac{i}{\pi} \cosh^{-1} \frac{P_M(v)}{2}\;.
\end{equation}
The large $v$ asymptotics is the expected $\gamma(v)\simeq \frac{iM}{\pi}\ln v$, with a convenient choice of branch cuts in the $v$ plane.
Moreover, 
\begin{equation}\label{dgamma_generic_poly}
\begin{split}
d\gamma(v)=\frac{i}{\pi}\frac{dP_M(v)}{\sqrt{P_M(v)^2-4}}\;.
\end{split}
\end{equation}
The previous exact expressions \eqref{gamma_generic_poly}-\eqref{dgamma_generic_poly} are to be compared with the naive expressions 
\begin{equation}\label{gamma_naive_generic_poly}
\gamma_n(v)=\frac{i}{\pi} \log P_M(v)
\end{equation}
and 
\begin{equation}\label{dgamma_naive_generic_poly}
\begin{split}
d\gamma_n(v)=\frac{i}{\pi}\frac{dP_M(v)}{P_M(v)}\;,
\end{split}
\end{equation}
which capture the semiclassical dynamics of the dual gauge theory but miss the nonperturbative effects encoded in the SW curve. The difference is that new branch cuts, related to the square root, appear in \eqref{gamma_generic_poly}. $d\gamma$ behaves more mildly than the naive $d\gamma_{n}$, which has simple poles at $v=v_i$: at most it has the singular behavior $d\gamma(v)\sim (v-v_{i\pm})^{-\frac{1}{2}}\, dv$ as $v\to v_{i\pm}$ if $P_M(v_{i\pm})=\pm 2$ and $ P_M'(v_{i\pm})\neq 0$ or branch points if $ P_M'(v_{i\pm})= 0$, otherwise it is analytic. The simple pole at infinity survives.




\subsection{Branch points and singularities of the SW curves}\label{subsec:branc_singul}

This section is devoted to the study of branch points and singularities of the SW curve \eqref{SW_curve_inf_cascade} or equivalently \eqref{SW_curve_inf_cascade_quiver}, and their relations with special values of the twisted sector complex scalar $\gamma$ in type IIB string theory.

Let us start with branch points, which correspond to values of $v$ such that $u_-(v)=u_+(v)$ up to the torus equivalence $u\sim u+1\sim u+\tau$. When this happens, the two branches of the M5 brane join; after reducing to type IIA string theory, this means that at those values of $v$ the two NS5 branes cross each other with no discontinuity of the periodic scalar on their worldvolume.
By construction, these branch points arise when $u\in\frac{\bbZ +\tau \bbZ}{2}$ \cite{Petrini:2001fk}.
From the point of view of the solution \eqref{gamma_generic_poly} for the twisted sector scalar potential, they correspond to values of $v$ such that $\gamma(v)\in \bbZ+\tau \bbZ$, by virtue of \eqref{gamma_from_u}. In particular, those values of $v$ are defined by the condition
\begin{equation}\label{branch_condition}
P_M(v)=\pm \left(q^\frac{n}{2}+q^{-\frac{n}{2}}\right)=2 \cosh\left[i \pi (n\tau+m)\right]\;,\qquad n,m\in\bbZ\;.
\end{equation} 
Notice that from the point of view of string theory these are locations where additional massless degrees of freedom might appear (in the type IIA picture tensionless strings arise where the two NS5 branes cross). Despite our poor knowledge of string and M theory in such conditions, in the present case gauge/string duality along with Seiberg-Witten theory allows us to identify the extra massless states whenever they appear.

The quickest way of seeing which of those branch points give rise to additional massless modes (monopoles, dyons or gauge bosons) is to look for singularities of the SW curve. They are most easily found using its form \eqref{SW_curve_inf_cascade}, and correspond to points $(t,v)$ such that $\tilde Q(t,P_M(v))=0$ and $d\tilde Q(t,P_M(v))=0$. 
The defining equation of the curve $\tilde Q(t,P_M(v))=0$ requires that, for some integer $h$, $t$ and $v$ are subject to the condition
\begin{equation}
q^h t+\frac{1}{q^h t}=P_M(v)\;.
\end{equation} 
Then the curve is singular if 
\begin{equation}\label{singul}
\lim_{K\to\infty} q^{K(K+1)}\,\left[q^h\left(1-\frac{1}{q^{2h}t^2}\right)dt-dP_M(v) \right]\prod_{\substack{
l=-K\\
l\neq h}}^K  \left(q^l t+\frac{1}{q^l t}-q^h t-\frac{1}{q^h t}\right)=0
\end{equation}
holds too.
There are two classes of possibilities, which we will name singularities of the first or of the second kind, depending on whether it is the first factor in \eqref{singul} or one of the factors in the infinite product that vanishes.

Singularities of the first kind are solutions of
\begin{equation}
\begin{cases}
q^h t =\pm 1\\
P_M(v) = \pm 2\\
P_M'(v)=0
\end{cases}
\end{equation}
and they may or may not exist depending on the form of the polynomial $P_M(v)$. They come from possible singularities of the (images under $t \mapsto q^h t$ of the) SW curve \eqref{SW_curve_glue_dimless} of pure $SU(M)$ SYM.
Correspondingly, at such values $v=v_*$ the twisted sector scalar $\gamma(v_*)\in \bbZ$, namely $c$ is integer and $b=0$.%
\footnote{Recall that a rescaling of the form $t \mapsto \alpha t$, in particular $t \mapsto q^h t$, leaves $\gamma$ invariant. }

Singularities of the second kind are solutions of
\begin{equation}
\begin{cases}
q^h t =\pm q^\frac{h-l}{2}\\
P_M(v) = \pm \left(q^\frac{h-l}{2}+q^{-\frac{h-l}{2}}\right)  =2\cosh\left[ i \pi \left((l-h)\tau+m\right) \right]
\end{cases}
\end{equation}
with $l\neq h$ and $l,m\in\bbZ$.
These singularities always exist, and correspondingly $\gamma(v)\in \bbZ+\tau\bbZ_0$, namely $c$ and $b$ are integer but $b\neq 0$.
They genuinely arise from the (infinitely) cascading nature of the RG flow of the vacua constructed; they are double branch points reminiscent of the baryonic root of $\calN=2$ SQCD, in the same spirit of \cite{Benini:2008ir} but with the important difference that now these branch points are by construction \emph{exactly} double for every choice of the polynomial $P_M(v)$, precisely like those at the baryonic root of $\calN=2$ SQCD. Mutually local massless monopoles arise, making the analogy to the baryonic root tighter.
 
Summarizing, branch points of the SW curve for the infinite cascade appear at values of $v$ such that $\gamma(v) \in \bbZ+\tau\bbZ$. Where $\gamma(v)\in \bbZ+\tau\bbZ_0$ we have exactly double branch points related to the strong coupling transitions in the cascade; the branch points where $\gamma(v)\in \bbZ$, instead, are generically not double, but can be made so by tuning zeros of the characteristic polynomial $P_M(v)$, precisely as for the curve of the $SU(M)$ pure gauge theory.


\section{Examples}\label{sec:examples}

In this section we apply the general method for finding exact solutions for the twisted fields developed in the previous section to some interesting cases: the $\bbZ_{2M}$-symmetric enhan\c con vacuum whose dual is approximated by the excised version of the solution of \cite{Bertolini:2000dk,Polchinski:2000mx} with $M$ smeared tensionless fractional branes at the enhan\c con ring, and one of the $M$ $\bbZ_2$-symmetric enhan\c conless vacua whose SW curves have genus zero \cite{Benini:2008ir} and which become the Klebanov-Strassler vacua after an infinite mass deformation.

\subsection{The $\bbZ_{2M}$--symmetric enhan\c con vacuum}\label{subsec:enhancon}

\begin{figure}[tn]
\centering \includegraphics[width=9cm]{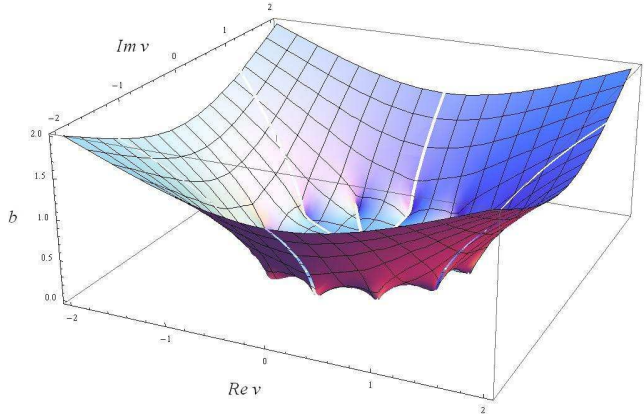}
\caption{\small $b$ field in the type IIB dual of the enhan\c con vacuum, for $M=6$ and $g_s=1$.  \label{fig:b-enhancon}}
\end{figure}

The $\bbZ_{2M}$-symmetric enhan\c con vacuum with an infinite cascade can be obtained by means of the previously explained embedding by taking $P_M(v)=v^M$. The exact twisted sector scalar field is
\begin{equation}\label{gamma_enhancon}
\gamma(v)=\frac{i}{\pi} \log \left[ \frac{v^M}{2}+ \sqrt{\left(\frac{v^M}{2}\right)^2 -1} \right]= \frac{i}{\pi} \cosh^{-1} \frac{v^M}{2}\;,
\end{equation}
whose imaginary part is plotted in Fig. \ref{fig:b-enhancon}.

We can approximate the previous exact result inside and outside the circle of radius $\rho_e\equiv 2^{1/M}$ and get to leading order
\begin{equation}\label{approx_enhancon}
\gamma(v)\simeq 
\begin{cases}
\frac{iM}{\pi}\log v +\calO(v^{-2M}) & \textrm{if}\quad |v|^{M}\gg 2 \\
-\frac{1}{2}+\frac{v^M}{2\pi}+\calO(v^{2M})  & \textrm{if}\quad |v|^{M}\ll 2 
\end{cases}\;,
\end{equation}
where $[\cdot]_-$ is the floor function and we worked in the $v$ plane with a suitable choice of branch cuts lying along the circle of radius $\rho_e$ and the real half-line $[2^{1/M},+\infty)$.

The approximation employed in \cite{Benini:2008ir} is recovered upon further neglecting the first corrections in the interior. As a bonus, we also get the average value of the potential in the interior. 

Note also, either using the exact result or the approximation, that the 5-brane charge enclosed in a disk of radius $\rho$ centered in the origin vanishes if $\rho<\rho_e$ and equals $2M$ if $\rho>\rho_e$, as expected.

The branch points of $\gamma$ \eqref{gamma_enhancon} at finite values of $v$ are the branch points of the SW curve of $SU(M)$ at the origin of the moduli space, namely the $2M$ roots of $v^{2M}=4=\rho_e^{2M}$, lying on the enhan\c con circle. They are simple branch points of the SW curve of the quiver theory. 
Exactly double branch points of the SW curve of the quiver theory correspond to the roots of $v^{2M}=4\left[\cosh(i\pi k\tau)\right]^2$, $k\in\bbZ_0$. 
Far from the enhan\c con region, where the logarithmic approximation can be applied, they lie at the intersections 
of circles (curves of integer $b$) with logarithmic spirals (curves of integer $c$).




\subsection{The vacuum with genus 0 hyperelliptic curve}\label{subsec:genus0}

\begin{figure}[tn]
\centering \includegraphics[width=9cm]{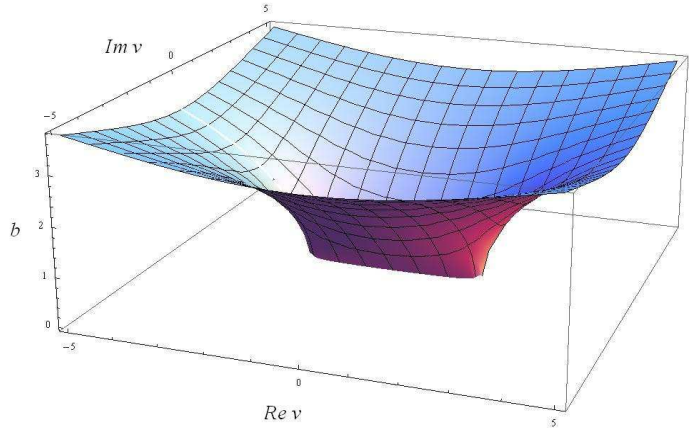}
\caption{\small $b$ field in the type IIB dual of the vacuum with genus zero SW curve, for $M=6$ and $g_s=1$.  \label{fig:b-enhanconless}}
\end{figure}

A $\bbZ_{2}$-symmetric enhan\c conless vacuum with an infinite cascade, whose Seiberg-Witten curve has genus zero, is obtained by taking $P_M(v)=2T_M\left( \frac{v}{2} \right)$, where $T_M$ is the $M$-th degree Chebyshev polynomial of the first kind, which satisfies $T_M(x)=\cos(M \cos^{-1}x)=\cosh(M \cosh^{-1}x)$. The asymptotics is
 $P_M(v)\approx v^M$ as $v\to\infty$.
The exact twisted sector scalar field \eqref{gamma_generic_poly} then becomes
\begin{equation}\label{gamma_genus_zero}
\gamma(v)= \frac{i}{\pi}\,\cosh^{-1} T_M\left(\frac{v}{2}\right)= \frac{iM}{\pi}\,\cosh^{-1} \frac{v}{2}\;,
\end{equation}
whose imaginary part is plotted in Fig. \ref{fig:b-enhanconless}.

The natural coordinates on the complex plane for the configuration under investigation are elliptic coordinates.
If we set $v=x+iy$, we can introduce new coordinates $\omega=\mu +i\nu$ according to
\begin{equation}\label{elliptic_coordinates}
v = 2 \cosh\omega \qquad \Longleftrightarrow  \qquad \begin{cases}
x = 2 \cosh\mu \cos\nu \\
y = 2 \sinh\mu \sin\nu 
\end{cases}
\end{equation}
so that
\begin{align}
\frac{x^2}{\cosh^2\mu}+\frac{y^2}{\sinh^2\mu} &=4   \label{ellipses}   \\
\frac{x^2}{\cos^2\nu}-\frac{y^2}{\sin^2\nu} &=4 \;. \label{hyperbolae}
\end{align}
We can take $\mu\in\bbR^+$, $\nu\in[-\pi,\pi]$. 
Equation \eqref{ellipses} tells us that curves of constant $\nu$ are ellipses and equation \eqref{hyperbolae} tells us that curves of constant $\mu$ are hyperbolae. The ellipses have semimajor axis $a= 2\cosh\mu$ and eccentricity $e= 1/\cosh \mu$; their foci are at $z=\pm 2$; at large $\mu$ they become more and more similar to circles.
Using these coordinates the twisted sector scalar takes the simple form
\begin{equation}\label{gamma_omega}
\gamma=i\,\frac{M}{\pi}\,\omega\;.
\end{equation}
Therefore curves of constant $b$ are those ellipses and curves of constant $c+C_0 b$ are those hyperbolae.

The branch points of $\gamma$ \eqref{gamma_genus_zero} at finite values of $v$ are the branch points of the SW curve of $SU(M)$ with $P_M(v)=2T_M\left( \frac{v}{2} \right)$, namely the points $v_m=2\cos\frac{\pi m}{M}$, $m\in\bbZ$, which lie on the degenerate ellipse \eqref{ellipses} with $\mu=0$, namely the segment $v\in[-2,2]$ on the real axis. They are branch points for the SW curves of $SU(M)$ and of the quiver theory, with the branch cuts attached to one another along this segment.
The other double branch points of the SW curve of the quiver theory lie at $v=2\cos\left( \frac{\pi}{M}(m+\tau n)\right)$, with $m\in\bbZ_0$ and $n\in\bbZ$. In the region of large $|n|/(g_sM)$, where the logarithmic approximation holds, they approximately lie at the intersection of circles and logarithmic spirals: $v\simeq \exp\left[\pi\left( \frac{n}{g_sM}-i\frac{m}{M} \right)\right]$. 

As expected, there is no true enhan\c con, in the sense that no region of the complex plane is enclosed by the innermost of the generalized enhan\c con loci (\emph{i.e.} the $b=0$ locus), because that degenerates to a segment.

The discussion of a closely related twisted field configuration, relevant to $\calN=2$ fractional D3 branes on orbifold fixed loci with cylindrical topology such as the $\bbZ_2$ orbifold of the deformed conifold, is relegated to the appendix.


\section{Dissolution of fractional branes into twisted fluxes}\label{sec:dissolution}

In the previous section we employed the embedding of the moduli space of the pure gauge theory into the Coulomb branch of the quiver gauge theory with an infinite cascade and the related construction of exact solutions for twisted fields in type IIB string theory explained in section \ref{sec:embedding} for studying configurations where all the eigenvalues of the adjoint scalars in the quiver gauge theory lie in nonperturbative regions. By the construction of section \ref{sec:embedding}, we can separate in \eqref{characteristic_polynomials} the eigenvalues of the two adjoint scalars into those of $P_M(v)$ and those related to the infinite cascade.
Correspondingly, in the analysis of subsection \ref{subsec:branc_singul} we also distinguished branch points of the SW curve of the quiver theory \eqref{SW_curve_inf_cascade_quiver} which are also branch points of the curve of the pure gauge theory \eqref{SW_curve_glue_dimless} (namely $n=0$ in eq. \eqref{branch_condition}) and those which are not ($n\neq 0$). The former ones may or may not be double according to the form of the polynomial $P_M(v)$, whereas the latter are exactly double by construction, in complete analogy to the branch points of $\calN=2$ SQCD at the baryonic root.

In this section we will concentrate on eigenvalues and branch points of the first class. The study of the exact dual twisted field configurations will allow us to understand the fate of fractional D3 branes in type IIB string theory at nonvanishing string coupling.

If the absolute values of one of such eigenvalues is much larger than $\rho_e=2^{1/M}\approx 1$ (in units of $\Lambda$), then the eigenvalue lies in a perturbative region of the $SU(M)$ theory where the semiclassical approximation is good and the nonperturbative splitting of the two related branch points is small. In the type IIA/M theory description, the D4 brane is inflated into a very thin M5 brane tube. Neglecting nonperturbative effects in the dual gauge theory, the type IIB description is in terms of a `localized' fractional D3 brane, leading to a simple pole of $d\gamma$ at its location.%

When instead the absolute value of the eigenvalue becomes comparable to or smaller than $1$, then the splitting between the two branch points becomes of the same order of magnitude as the separation between different pairs of branch points. In the type IIA/M theory description, the D4 brane is inflated into a fat M5 brane tube, that might even touch other fat tubes in more singular situations. In the type IIB description, one could think that the fractional D3 brane develops a wavefunction which is spread over a region of the size of the splitting. 

In the previous section we studied two configurations dual to nonperturbative points of the moduli space of the $SU(M)$ theory, the enhan\c con vacuum and the enhan\c conless vacuum whose hyperelliptic curve is a sphere. The exact form of twisted fields in such cases is already quite instructive for our purposes.

The naive enhan\c con ring solution proposed in \cite{Benini:2008ir} involved $M$ tensionless fractional D3 branes smeared at a ring of radius $\rho_e$ (the enhan\c con ring). In that approximation, the tensionless smeared fractional branes were needed as sources accounting for the discontinuity of the enclosed D5 brane charge at the ring. Instead, our exact solution \eqref{gamma_enhancon} involves twisted fields only, with no need of smeared tensionless fractional brane sources. Indeed, the solution for $\gamma$ is holomorphic everywhere along the ring, except at the $2M$ branch points of the square root,%
\footnote{Of course $\gamma$ is everywhere well defined on its Riemann surface.}
 and the discontinuity of the D5 brane charge is simply provided by the $M$ disjoint branch cuts joining them, along with the $M$ branch cuts lying between one branch point in each pair and the point at infinity. 

Therefore we see that at nonvanishing string coupling those fractional D3 branes, rather than becoming tensionless and uniformly smeared along the ring, completely dissolve into twisted fluxes. The same statement holds for the enhan\c conless configuration of subsection \ref{subsec:genus0}, for which all the branch cuts join to become a single one. After the transmutation, the D5 brane charges of the fractional D3 branes are provided by the monodromies of $\gamma$ around the nontrivial loops of the Riemann surface $\gamma$ is defined on. We will detail this assertion more in the following. 

Surprisingly enough from the naive viewpoint of type IIB string theory, we will see in the remainder of this section that the transmutation into twisted fields occurs as well for naively tensionful fractional branes, related to eigenvalues lying in a perturbative region. This dissolution also addresses two interrelated puzzles concerning localized tensionful fractional D3 branes. The resolution of the puzzles is provided by gauge theory instantonic effects encoded in the SW curve. Those effects might look negligible in the gauge theory in regimes of small coupling; in spite of that, regardless of how small they are, they turn out to be always crucial for solving the two puzzles on the dual string side, that we now explain.

\subsection{Puzzles with fractional D3 branes}

Consider $M$ coincident fractional D3 branes, located conventionally at $v=0$, and let them backreact.
Naively, they contribute a factor $\frac{iM}{\pi}\log v$ to the twisted sector complex scalar potential $\gamma$.
The twisted fluxes sourced by the fractional branes carry a D3 brane charge because of the modified Bianchi identity/equation of motion 
\begin{equation}\label{eq_F5}
dF_5=-H_3\wedge F_3+\mathrm{sources}=\frac{i}{2}g_s (4\pi^2\alpha')^2 \,d\gamma\wedge\overline{d\gamma}\wedge\omega_2\wedge\omega_2 +\mathrm{sources}\;,
\end{equation}
where $\omega_2$ is a closed antiselfdual $(1,1)$ form with delta-function support on the orbifold plane, normalized as $\int_\calC \omega_2=1$, which satisfies $\int_{ALE}\omega_2\wedge \omega_2=-\frac{1}{2}$, and $F_5$ the gauge invariant improved RR 5-form field strength.
The fractional branes themselves, if tensionful, also contribute to \eqref{eq_F5} via their D3 brane charge, which in turn depends on the value of $b$ at their location. Both those contributions (twisted fluxes and D3 brane charge) source $F_5$.

We now face a first problem: considering a shell $S$ with $S^5/\bbZ_2$ boundaries of outer radius $R_o$ and inner radius $R_i$ centered in the position of the fractional D3 branes, the contribution of the twisted fields generated by the fractional branes to the D3 charge is
\begin{equation}
\Delta Q_3^{fluxes} (S)\equiv -\frac{1}{(4\pi^2\alpha')^2}\int_{S}F_5=\frac{g_s M^2}{\pi}\log\frac{R_o}{R_i}=M[b(R_o)-b(R_i)]\;,
\end{equation}
which diverges to $+\infty$ as we send $R_i\to 0$ keeping $R_o$ fixed. At the same time, the charge carried by the fractional D3 branes after the backreaction is formally $M b(0)$, which equals $-\infty$ after the backreaction is taken into account. These two divergences constitute the first problem. The guts of the problem are the fact that the fractional branes could not actually be BPS, if they really had negative D3 brane charge. One could object that each fractional brane should not backreact on itself, but that does not solve the problem: imagining to separate slightly the $M$ branes, one could still compute the effect on a single brane exerted by the other branes; then, as they approach each other, the D3 brane charge of each single brane decreases, at some point becoming negative and eventually diverging to $-\infty$ in the limit where it meets one of the other branes.

Note also that if we regularized the D3 brane charge carried by the fractional D3 branes by substituting $M b(0)$ with $M b(R_i)$ and added the two contributions without caring about divergences in the $R_i\to 0$ limit, we would then get a sensible answer, which is precisely the expected total D3 brane charge. 

The problem is analogous to the divergence of the self-energy of an electron in classical physics; there one has to assign by hand a classical size to the electron in order to avoid the divergence. If this size is instead sent to zero, then the electrostatic energy diverges. A usual prescription in classical electromagnetism is to substitute the electron with a conducting sphere whose radius, called the classical radius of the electron, is such that the whole mass of the electron is provided by the electrostatic potential energy of its field. 

A similar prescription could in principle be applied to the naive solutions under consideration, by promoting each fractional brane to a `conducting' BPS extended object. By `conducting' and BPS here we mean that $b$ is constant and nonnegative at the surface of this object. The previous demand does not uniquely define the surface, because of the arbitrariness of the boundary value of $b$. One might be tempted to fix the boundary value of $b$ so that the D3 brane charge of the fractional brane is the smallest allowed by supersymmetry, namely zero D3 brane charge. However, this prescription is clearly \emph{ad hoc} and does not explain if and how the problem is solved by string theory.

Note also that the divergence problem can be by-passed if the fractional branes are continuously distributed. That was done in \cite{Benini:2008ir} both for tensionful antifractional branes providing an ultraviolet cutoff to the cascading RG flow and for tensionless fractional branes lying at the enhan\c con ring in the infrared region of the dual field theory. If that smearing could perhaps be physically motivated for tensionless fractional branes, because SW theory teaches us that in some sense their wavefunctions are spread over a region of comparable size to the separation between different branes, it was nothing but a trick in the case of tensionful branes, whose wavefunctions are on the contrary very localized. The trick was necessary since it was not known how to deal with isolated fractional branes.

The second and related puzzle arises when trying to compute the warp factor. The contribution of twisted fluxes to the warp factor is computed via the following formula:
\begin{equation}\label{warp}
Z_{fl}(y,\vec{x})=4\pi \alpha'^2 g_s^2 \,\frac{i}{2} \int \frac{d\gamma(v)\wedge \overline{d\gamma(v)}}{\left[\vec{x}^2+|y-v|^2 \right]^2}\;,
\end{equation}
where $y\in \bbC$ is a coordinate along the directions parallel to the orbifold plane and $\vec{x}$ is a coordinate on the covering space $\bbR^4$ of the orbifold.
For the very same reason underlying the divergences in the D3 brane charge -- after all, for BPS objects charge and mass are related -- if there are tensionful fractional brane sources the integral \eqref{warp} diverges \emph{everywhere} \cite{Bertolini:2000dk}: the integrand $d\gamma\wedge\overline{d\gamma}$ in \eqref{warp} close to the sources is not integrable. On the other hand, there would be an additional contribution from the (negative infinite) localized D3 brane charges of the fractional branes. After formal (and to some extent arbitrary) regularization and subtraction similar to those mentioned for the D3 brane charge, one can see that the two divergences cancel.
Again, the smearing trick was used in \cite{Benini:2008ir} to by-pass the problem.


\subsection{Resolution of the puzzles by transmutation of fractional branes into twisted fields}

Exploiting SW theory and dualities, in section \ref{sec:embedding} we managed to find the exact twisted fields configurations \eqref{gamma_generic_poly} in type IIB string duals of a class of infinitely cascading vacua.
With those solutions at hand, in the remainder of this section we will show how type IIB string theory at nonzero coupling resolves the issues of divergences explained in the previous subsection, by complete transmutation of the fractional D3 branes into twisted fields. We will find that the D3 brane charge enclosed in any finite region is finite, and that the contribution of the twisted fluxes to the metric is finite too.%
\footnote{Except for the expected curvature singularity which is met when approaching the orbifold plane where localized field strengths have support.}
The mechanism is general and applies both to fractional branes which are naively tensionless (eigenvalues in highly nonperturbative regions for the gauge theory) and to fractional branes which are naively tensionful (eigenvalues in perturbative regions for the gauge theory). 

\begin{figure}[tn]
\centering \includegraphics[width=10cm]{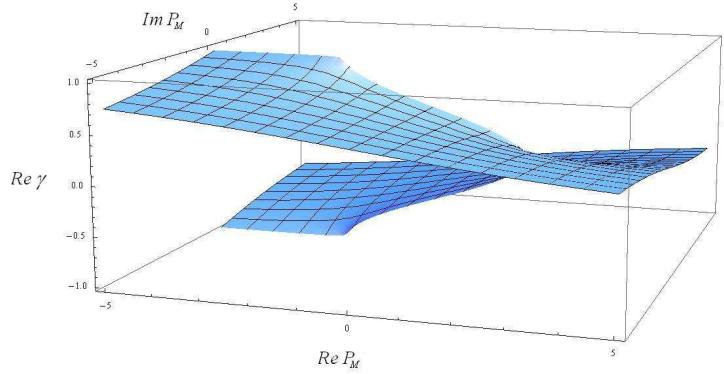}
\caption{\small Real part of $\gamma$ field in the complex plane parameterized by $P_M$.  The imaginary part is shown in Fig. \ref{fig:b-enhanconless}, up to an overall factor of $M$. \label{fig:Re-gamma-P}}
\end{figure}

Recall the exact formula \eqref{gamma_generic_poly} for the twisted field $\gamma$ and consider its dependence on $P_M$. In the $P_M$ complex plane, $\gamma$ has a branch cut of the square root, that we conveniently take along the real segment $P_M\in[-2,2]$, and a branch cut of the logarithm, that we can take on the negative real axis along the half-line $P_M\in(-\infty,-2]$, see Fig. \ref{fig:Re-gamma-P}. Consider now a counterclockwise loop that winds once around the branch cut of the square root: picking the positive determination of the square root of 1, it is easy to compute $\oint d\gamma=-2$. We can parametrize the branch cut by setting $P_M=2\cos\beta$, with $\beta\in\bbR$ and $\beta\in[-\pi,\pi]$; hence  $\gamma=-\beta/\pi$ along the cut, and we wind around it counterclockwise from $P_M=-2$ to itself as $\beta$ varies between $-\pi$ and $\pi$. We observe that $b=0$ along the curve joining $P_M=-2$ and $P_M=2$, and $c$ varies from $1$ to $0$ as $P_M$ moves along the segment rightward from below and then from $0$ to $-1$ as $P_M$ moves along the segment leftward from above.

\begin{figure}
\centering \includegraphics[width=12cm]{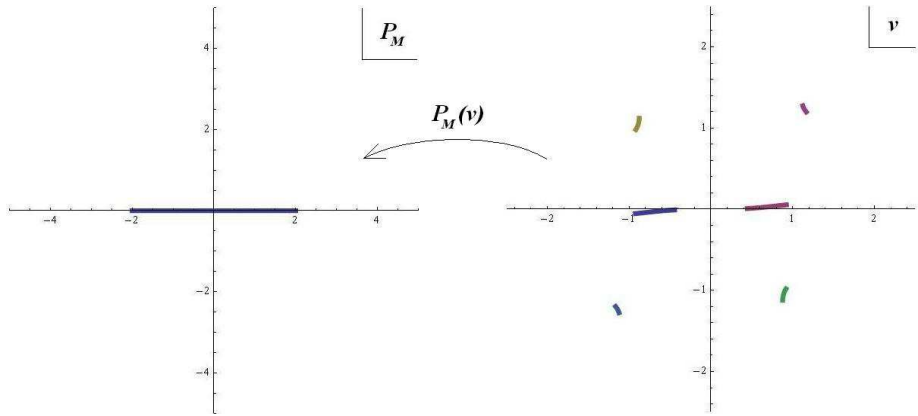}
\caption{\small Example of branch curves in the $v$ plane for $M$=6. They are the inverse image of the segment $[-2,2]$ under the polynomial function $P_M(v)$. \label{fig:branchcurves-map}}
\end{figure}

If we switch from $P_M$ to $v$, recalling that $P_M(v)$ is a degree $M$ polynomial in $v$ we get $M$ roots for $v$ as a function of $P_M$. For generic polynomials, the roots are all different and the curve joining $P_M=-2$ and $P_M=2$ along the real $P_M$ axis gives rise to $M$ curves joining pairs of branch points of the SW curve of $SU(M)$ in the $v$ plane, as shown in Figure \ref{fig:branchcurves-map}. We will name these curves `branch curves'. Along these branch curves $b=0$, and encircling any of them counterclockwise the monodromy of $c$ is $-2$. Each of these branch curves is the remnant of a fractional D3 brane, as we see from the fact that the monodromy leads to the D5 brane charge of a fractional D3 brane. Indeed the 5-brane charge is 
\begin{equation}
Q_5 = -\frac{1}{4\pi^2\alpha'}\int_{\calC\times \calL} F_3=- \mathrm{Re} \oint_\calL d\gamma= -\oint_\calL d\gamma\;,
\end{equation} 
where $\calL$ is a loop in the $v$ plane, and each fractional brane has charge $2$, because of the self-intersection number of the exceptional 2-sphere $\calC$.

We remark that no fractional brane sources have to be added: fractional branes have transmuted into twisted fluxes. This is different from having tensionless smeared fractional branes, which appear as sources of D5 brane charge in the equation of motion of $\gamma$. This phenomenon is the T-dual mechanism of type IIA D4 branes inflating into M5 brane tubes or the type IIB string dual of the splitting of branch points in the SW curve of the gauge theory.
The result that all fractional branes transmute into twisted fluxes can be understood as follows: in the M theory description all the information is encoded in the holomorphic embedding of an M5 brane, and under the duality between M theory and type IIB string theory the fivebrane embedding translates into the twisted sector complex scalar potential. 
The absence of sources follows from the smoothness of the embedding.

With the correct solution at hand, there is no problem with D3 brane charge anymore: the total charge contained in any finite region is finite and positive, and there are no localized sources of D3 brane charge, unless harmless regular D3 branes are added. 
Indeed, the total D3 brane charge contained inside a 6-dimensional compact domain $\calS\in \bbC\times\bbC^2/\bbZ_2$ intersecting the orbifold plane on a 2-dimensional domain $\calD$  is \cite{Benini:2008ir} 
\begin{equation}\label{total_D3_charge}
Q_3(\calS)=-\frac{1}{(4\pi^2\alpha')^2}\int_{\partial\calS} F_5=\frac{1}{2}\int_\calD dc\wedge db=-\frac{1}{2}\int_{\partial\calD}b \,dc \;,
\end{equation}
where the branch curves have to be included in $\partial\calD$. In the situation that we are considering, $b=0$ along those curves and therefore there is no such contribution to \eqref{total_D3_charge}. In the next section we will enjoy the more general possibility where both fractional and antifractional branes are present.%
\footnote{We adhere to a common abuse of terminology, calling antifractional D3 brane the brane that together with a fractional D3 brane can form a regular D3 brane as a marginal bound state.}
 We will see that in such a case antifractional branes also transmute into fluxes, but along the corresponding branch curves $b=1$ and the monodromy of $\gamma$ is $+2$, so that their contribution to \eqref{total_D3_charge} provides the D3 brane charge of fractional/antifractional D3 brane pairs. 

Tuning the polynomials, it is possible to make some of the branch points of the $SU(M)$ curve collide, so that two or more of the branch curves join. In such situations we are in highly nonperturbative regimes where extra massless matter degrees of freedom (mutually local or nonlocal) appear. The most singular case is the enhan\c conless vacuum whose SW curve has genus zero, for which the branch curves all merge into a single one on the real $v$ segment between $-2$ and $2$. 

Finally, we can see how the problem with the warp factor disappears as well, once the correct solution for twisted fields is considered in \eqref{warp}, and taking into account that no additional contributions exist, because of the absence of sources with nonvanishing D3 brane charge (excluding regular D3 branes).
Recall that the curvature diverges approaching the orbifold plane where twisted fluxes have support. The metric of the supergravity plus massless twisted fields solution can not be trusted in that region, whereas it can be trusted far from it, provided the warp factor is well defined. We will therefore concentrate on locations $\vec{x}\neq \vec{0}$. Potentially dangerous integration regions in \eqref{warp} then correspond to singularities of $d\gamma(v)$. 
The problem with the naive solution \eqref{dgamma_naive_generic_poly} was that the integral \eqref{warp} does not converge around the zeros of the polynomial $P_M(v)$, irrespective of $y$ and $\vec{x}$.
With the exact solution \eqref{dgamma_generic_poly}, potentially dangerous integration regions are those surrounding the branch points of $\sqrt{P_M(v)^2-4}$. Let us expand $P_M(v)$ about one of those points, $v_0$:
\begin{equation}
P_M(v)=\pm \left[2+ a_l (v-v_0)^l+\calO\left((v-v_0)^{l+1}\right)\right]\;,\qquad 1\leq l\leq M\;.
\end{equation}
Then 
\begin{equation}
d\gamma(v) \simeq i \,\frac{l \sqrt{a_l}}{2\pi}\,(v-v_0)^{\frac{l}{2}-1} \,dv
\end{equation}
and therefore the contribution to \eqref{warp} coming from integration over a small neighborhood of $v_0$ with radius $\eps\ll 1$ is 
\begin{equation}\label{delta_warp}
\begin{split}
\delta Z(y,\vec{x})& \simeq \frac{ 4\pi \alpha'^2 g_s^2}{\left[\vec{x}^2+|y-v_0|^2 \right]^2}\,\frac{i}{2} \int_{|w|<\eps} \hspace{-13pt} d w \wedge \overline{d w} \; \,\frac{l^2 |a_l|}{(2\pi)^2} |w|^{l-2} = \frac{2\alpha'^2 g_s^2}{\left[\vec{x}^2+|y-v_0|^2 \right]^2}\,l |a_l|\, \eps^l\;,
\end{split}
\end{equation}
which is finite. Therefore the warp factor is finite for any $\vec{x}\neq \vec{0}$.

The previous analysis holds for the branch points related to $SU(M)$ SYM, which are not necessarily double. We will argue at the end of the next section that the same results are also valid for the exactly double branch points which are related to the infinite cascade.


\section{Remarks on duals of generic points of the moduli space of the quiver theory}\label{sec:HvsC}

So far we have studied the embedding of the moduli space of the $\calN=2$ pure gauge theory into the infinite-dimensional Coulomb branch of the quiver gauge theory on branes at the $\bbC\times\bbC^2/\bbZ_2$ orbifold with an infinite cascade, and the related type IIB dual solutions in the so called C-picture. In the situations we have studied, an infinite cascade was required in order for all the branch points to be double, except for the $2M$ of them arising from $SU(M)$ pure gauge dynamics.

In this section we will make several qualitative remarks about the correct solution for twisted fields in generic vacua of the quiver gauge theory, although we will not be able to provide an explicit analytic expression for $\gamma$: we will first consider CFT's with finite rank gauge groups in the ultraviolet like in \cite{Kachru:1998ys}, and see that the solution leads us to the so called H-picture of \cite{Benini:2008ir}, where the NSNS twisted sector potential $b$ is bounded between two adjacent integers, that we will choose to be 0 and 1. Such a picture is always valid, even in the presence of cascades with strong coupling transitions and enhan\c con bearings.
Then we will comment under which (very restricted) circumstances the C-picture arises as an equivalent description of the same physics, and motivate its validity in the construction of section \ref{sec:embedding}. 

We can anticipate that the reason why the C-picture is not always valid, even in the presence of cascades with strong coupling transitions and enhan\c con bearings, is that generically fractional branes located in regions with strong dual gauge coupling do not form domain walls for the twisted fields, as opposed to what appeared in the smeared ring approximation used in \cite{Benini:2008ir}. 

The SW curve for the $\calN=2$ $SU(N)\times SU(N)\times U(1)$ quiver gauge theory describing the low energy dynamics on $N$ regular D3 branes at the $A_1$ singularity is \cite{Witten:1997sc,Ennes:1999fb,Petrini:2001fk}
\begin{equation}\label{SW_quiver}
\frac{R_N(v)}{S_N(v)}=\frac{\theta_2(2u|2\tau)}{\theta_3(2u|2\tau)}\equiv g(u|\tau)\;,
\end{equation}
where $R_N(v)$ and $S_N(v)$ are the degree $N$ characteristic polynomials of the two adjoint scalar fields.%
\footnote{The complexified gauge couplings of the two groups are chosen to be equal in the ultraviolet.}
Recalling \eqref{torus_identification_t}, we will make use of the following formulae:
\begin{equation}
g(u|\tau)= q^{1/4}(t+t^{-1})\prod_{j=1}^\infty \frac{(1+t^2 q^{2j})(1+t^{-2} q^{2j})}{(1+t^2 q^{2j-1})(1+t^{-2} q^{2j-1})}
\end{equation}
and 
\begin{equation}
g(u+\frac{\tau}{2}|\tau)=g(u|\tau)^{-1}\;,\qquad\qquad g(u+\frac{1}{2}|\tau)=-g(u|\tau)\;.
\end{equation}
The SW curve \eqref{SW_quiver} appears in M theory as an M5 brane embedding in $\bbR^2\times T^2$, which is holomorphic with respect to the complex coordinates $v$ and $u$ in $\bbR^2$ and $T^2$ respectively. It can be equivalently viewed as a double cover of the plane (giving the locations of two NS5 branes as functions of $v$ after reducing to type IIA string theory) or an $N$-tuple cover of the torus (giving the locations of $N$ pairs of suspended D4 branes as functions of $u$ or $t$).
Note also that $g(-u|\tau)=g(u|\tau)$, which makes the double cover manifest. 

Given a point on the Coulomb branch of the quiver gauge theory, we can in principle invert \eqref{SW_quiver} and find the locations of the two NS5 branes $t_-(v)$ and $t_+(v)=t_-(v)^{-1}$. Then the holomorphic twisted scalar potential $\gamma(v)=c(v)+\tau\, b(v)$ in type IIB string theory is still given by \eqref{gamma_from_u}-\eqref{u_from_t}, with the difference that now $t_\pm(v)$ obey \eqref{SW_quiver}.

It turns out to be convenient to define again `branch curves' in the $v$ plane, as loci of integer $b$. They connect branch points of the SW curve \eqref{SW_quiver}, where $u=0,\frac{1}{2},\frac{\tau}{2},\frac{1+\tau}{2}$ modulo periodicities of the torus \cite{Petrini:2001fk}, and pass either through zeros of $R_N(v)$ or of $S_N(v)$. At those branch points $g(u|\tau)$ takes the values $g_0(q),-g_0(q),g_0(q)^{-1},-g_0(q)^{-1}$ respectively, where 
\begin{equation}
g_0(q)\equiv g(0|\tau)=2q^{1/4}\prod_{j=1}^\infty \frac{(1+ q^{2j})^2}{(1+ q^{2j-1})^2}\;.
\end{equation}
Each branch curve with even $b$ passes through a zero of $R_N(v)$, where $c$ is half-integer, and connects a branch point where $c$ is odd (and $g=-g_0(q)$) to a branch point where $c$ is even (and $g=g_0(q)$). 
Each branch curve with odd $b$ passes through a zero of $S_N(v)$, where $c$ is half-integer, and connects a branch point where $c$ is odd (and $g=-g_0(q)^{-1}$) to a branch point where $c$ is even (and $g=g_0(q)^{-1}$).
The branch curves are remnants of fractional and antifractional D3 branes, after they transmute into twisted fluxes.

Since $b(v)$ is a continuous function, if there exist two branch curves in the $v$ plane where, say, $b=0$ and $b=2$ respectively, then every continuous path connecting the two of them must cross a branch curve where $b=1$. Hence we conclude that the H-picture, where $b$ is bounded between $0$ and $1$ (by convention), is singled out, unless some branch curves join in such a way that they form a domain wall in the $v$ plane. In the H-picture each of the $N$ branch curves with $b=0$ gives a monodromy $\Delta c=-2$ and is the remnant of a fractional brane, whereas each of the $N$ branch curves with $b=1$ gives a monodromy $\Delta c=2$ and is the remnant of an antifractional brane.

In the approximation employed in \cite{Benini:2008ir} to describe cascading RG flows, smeared tensionless (anti)fractional D3 branes bounded regions of running $\gamma$ and so called enhan\c con plasma regions, namely regions (possibly of vanishing area) of constant integer $b$, hence forming smooth domain walls in the $v$ plane. Twisted fields could be solved for in each domain separately (the problem is a Dirichlet problem supplemented by the D5 brane charge quantization requirement), and then the solutions were glued by requiring continuity of $b$. In that approximation we had the freedom of reversing the sign of twisted field strengths in any region bounded by enhan\c con plasma regions, provided that at the same time we reversed as well the interpretation of the smeared sources at the boundaries, interchanging tensionless fractional and antifractional branes.%
\footnote{In the T-dual description, such freedom amounted to that of interpreting the same brane configuration, with smeared D4 branes of no extension in the $x^6$ direction, in terms of two intersecting fivebranes or rather two fivebranes touching each other in the enhan\c con plasma regions.}
Such freedom allowed us, in the description of rotationally symmetric configurations like those more frequently studied in the literature, to switch from the H-picture, where $b$ is always bounded between 0 and 1 and which appeared to be more natural in $\calN=2$ settings, to the C-picture, where $b$ is instead a monotonic function of the radial coordinate, like in the gravity duals of cascades of Seiberg dualities. More details can be found in \cite{Benini:2008ir}.

If instead we consider the correct solutions for twisted fields, smooth domain walls appear much more rarely. This can be easily understood by recalling that already in the $\bbZ_{2M}$-symmetric enhan\c con vacuum of section \ref{subsec:enhancon} the branch points at the enhan\c con ring in the IR are not double. Each branch curve is the union of two radial segments joining the origin with two adjacent branch points lying on the enhan\c con ring. These curves do not form a domain wall at the enhan\c con ring and indeed there is no inner region where $\gamma$ is constant.

\begin{figure}[tn]
\centering \includegraphics[width=3cm]{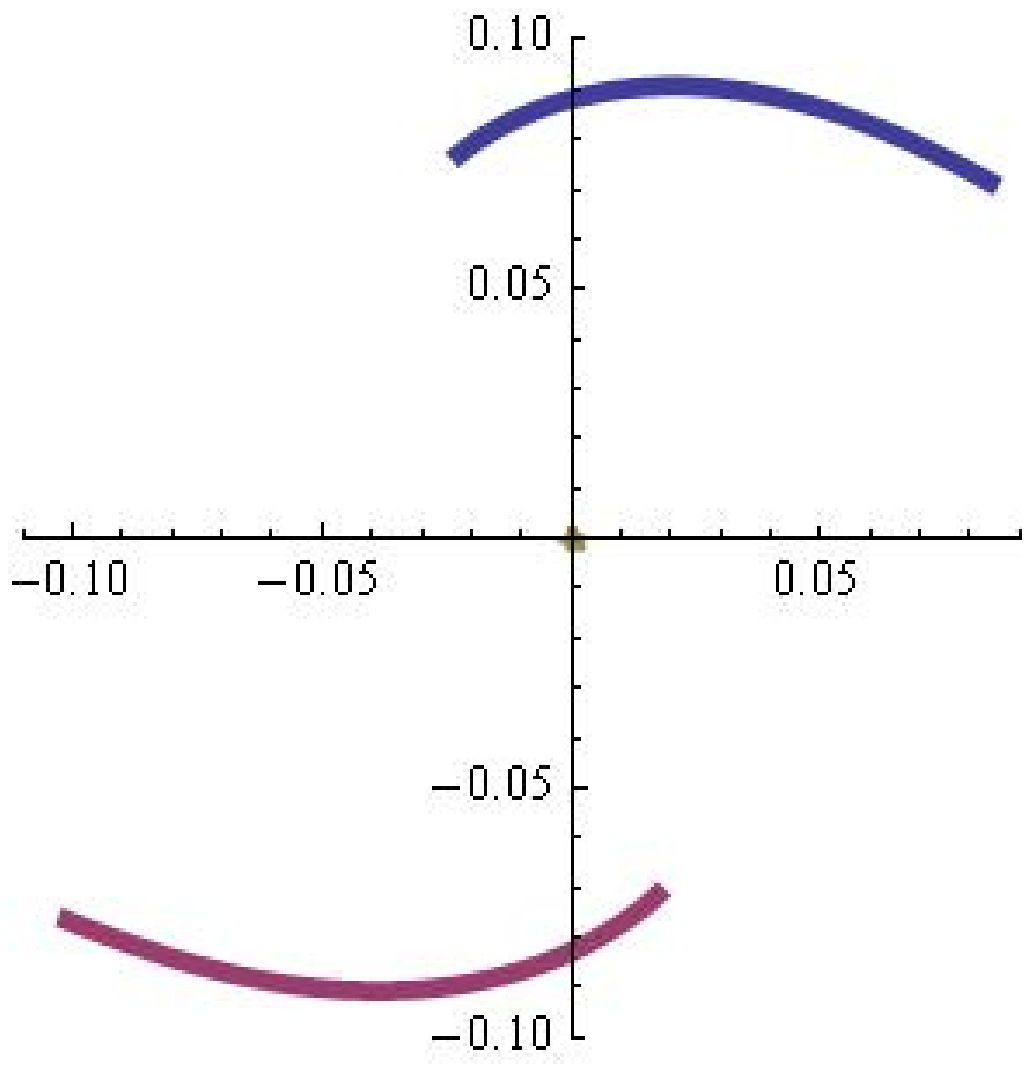}
 \includegraphics[width=3cm]{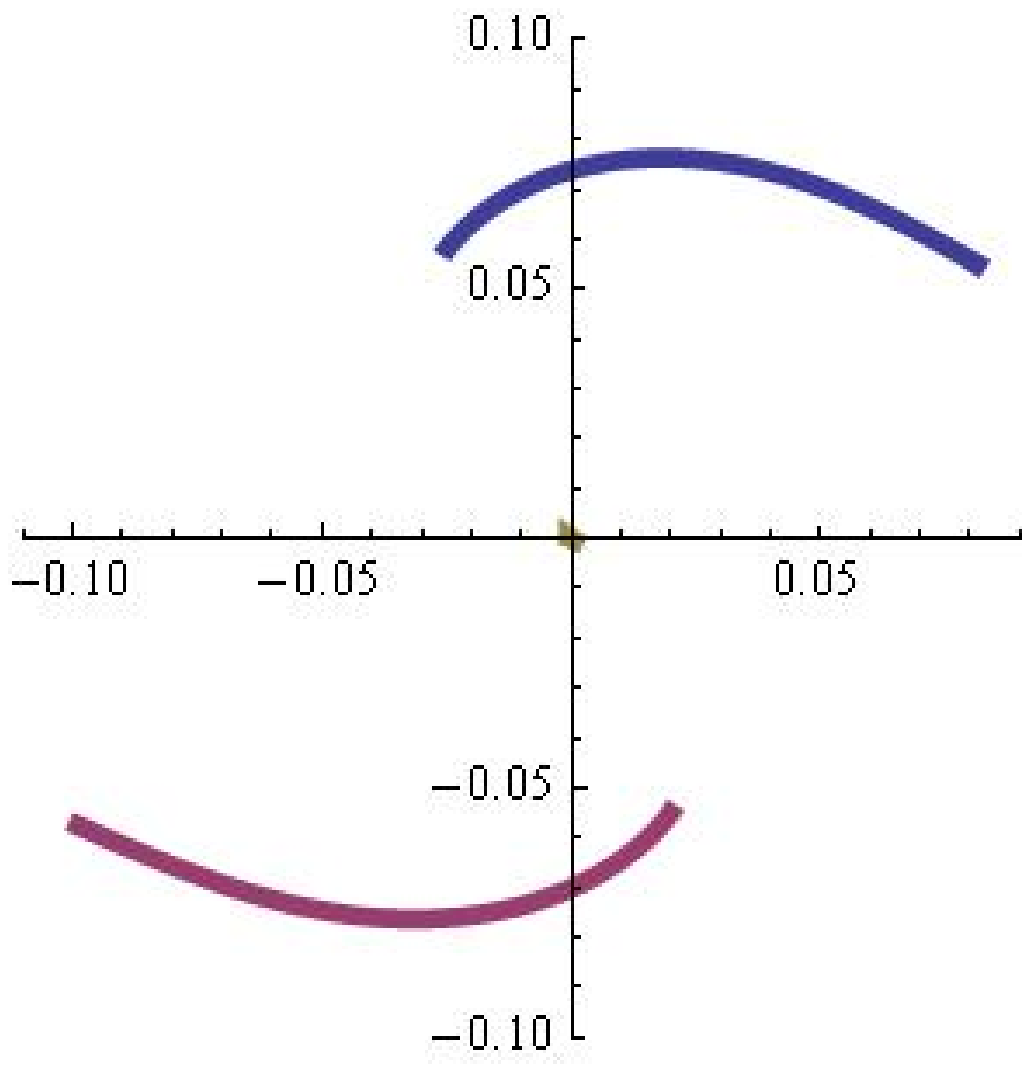}
 \includegraphics[width=3cm]{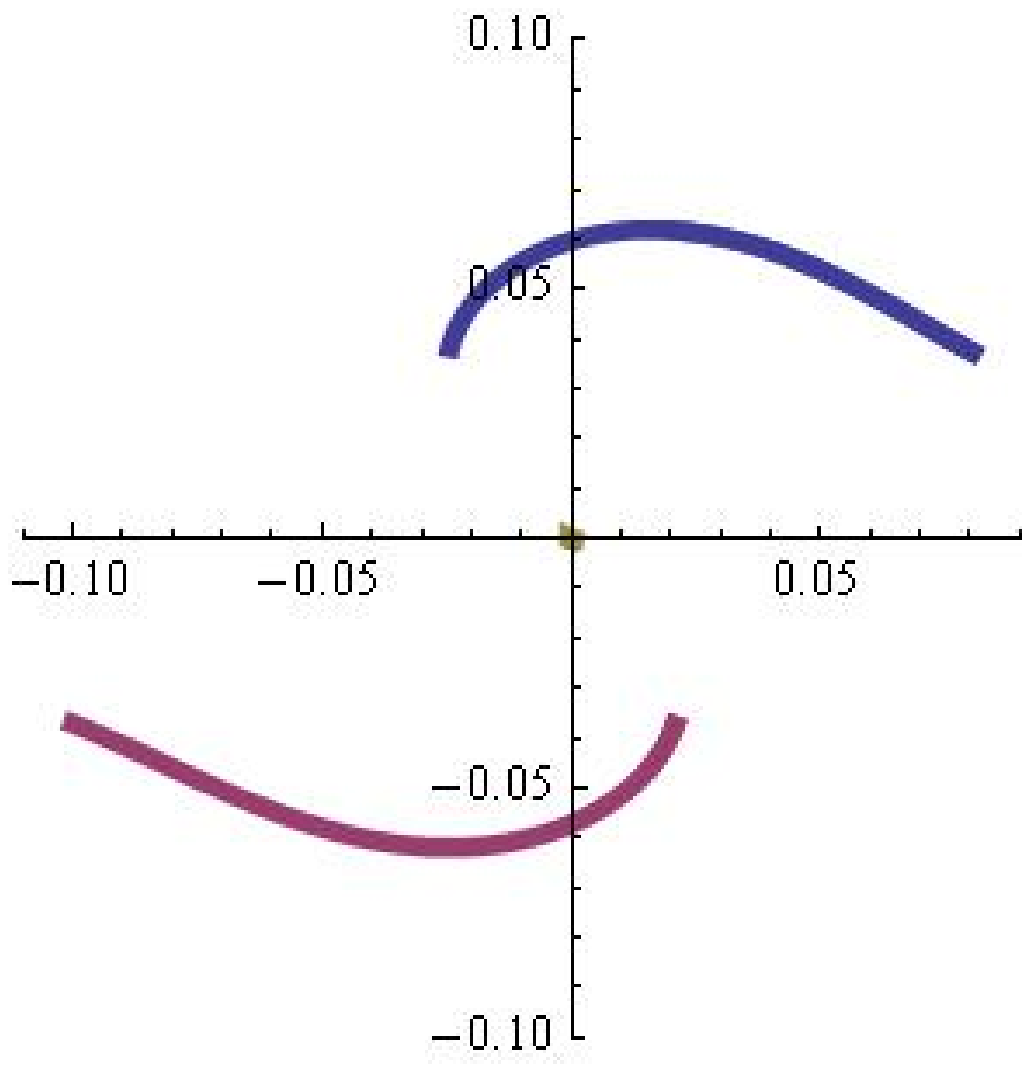}
 \includegraphics[width=3cm]{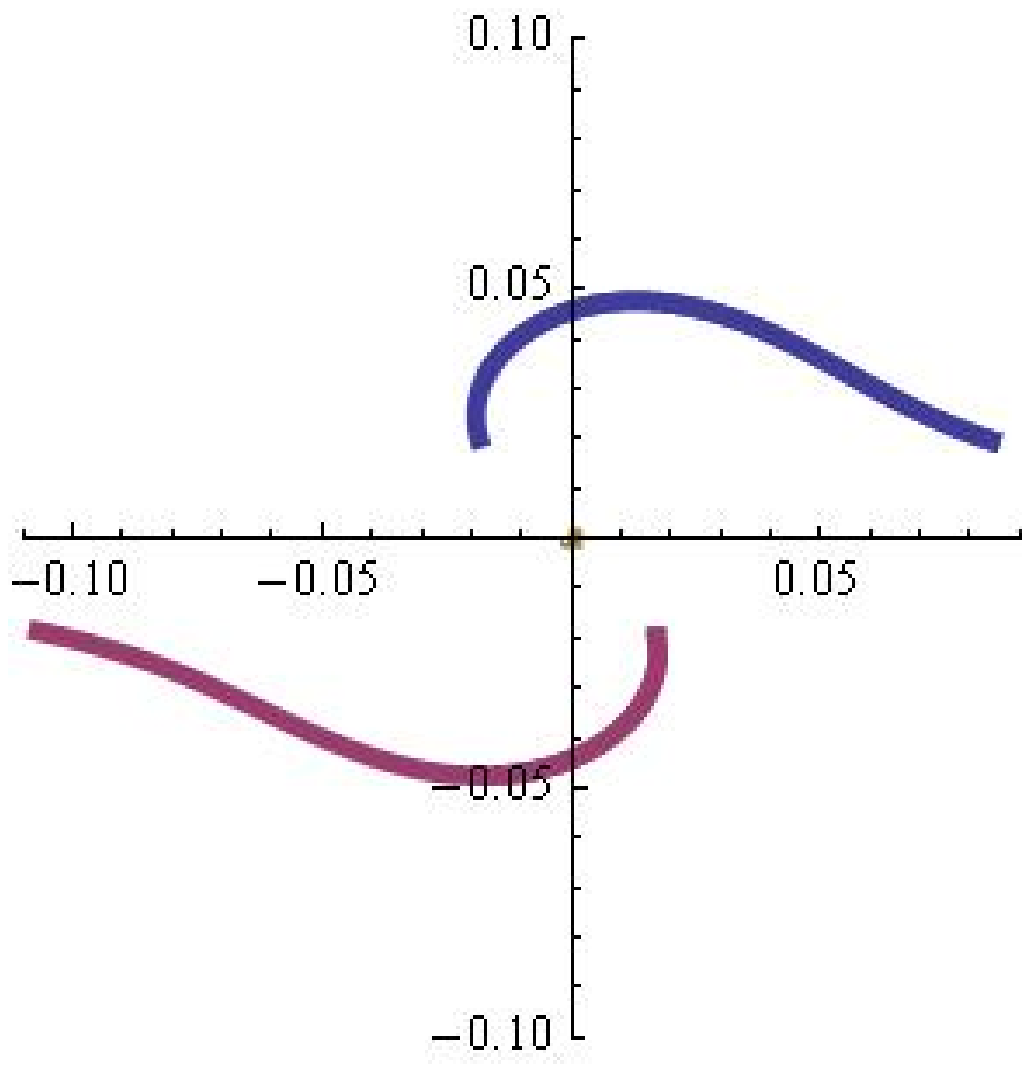}
 \includegraphics[width=3cm]{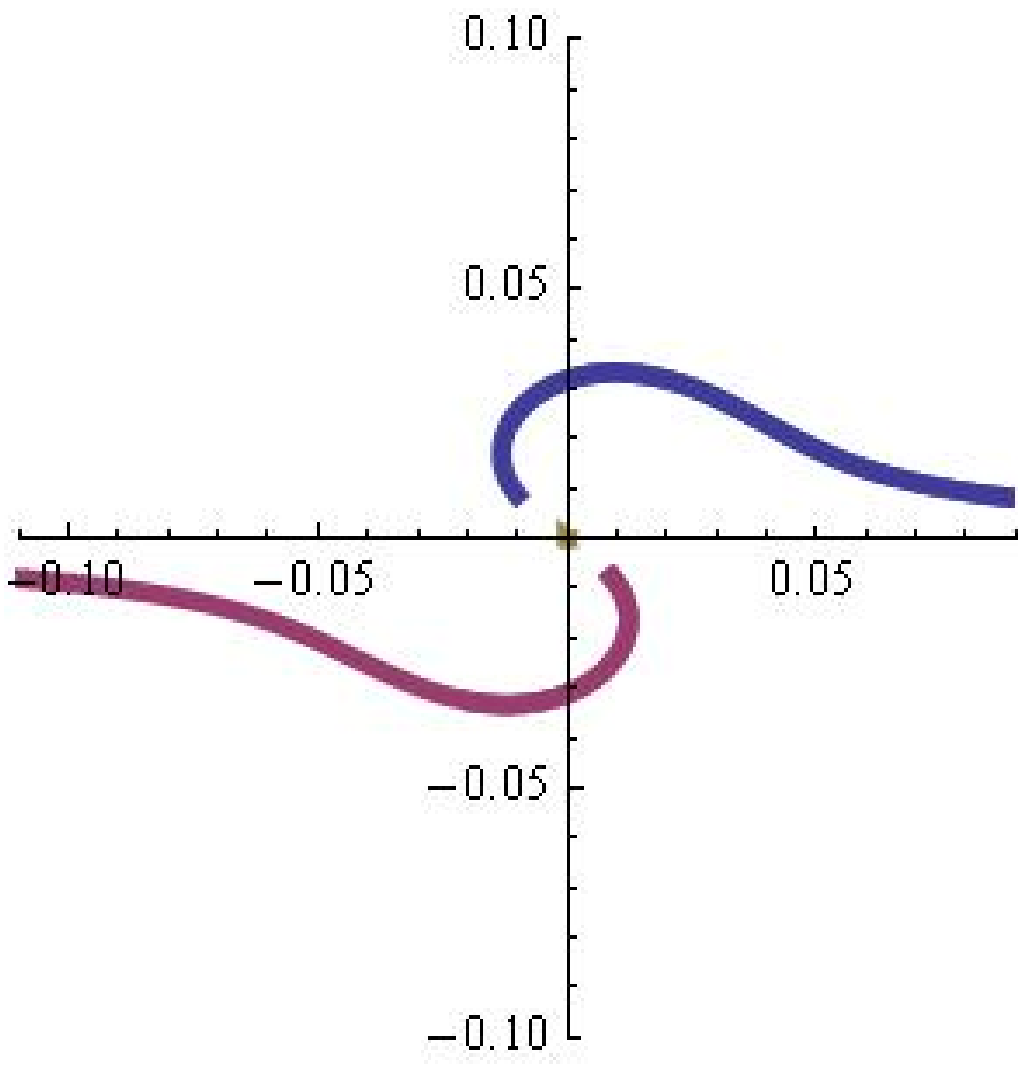}
\caption{\small Branch curves related to the enhan\c con bearing (in magenta and blue) for a generic choice of the phase of $\Phi$, shown in the $\xi$ plane, with $Z,q>0$. The branch curve passing through the origin is almost invisible due to the exponential hierarchy, and the one passing through $Z$ is out of the plot. $|\Phi|$ decreases from the left to the right.  \label{fig:branch_curves_bear_g}}
\end{figure}

It is instructive to observe what happens to branch curves in the procedure of forming the enhan\c con bearings of \cite{Benini:2008ir}, namely regions of enhan\c con plasma bounded by two concentric enhan\c con rings. They describe energy ranges in the RG flow of the dual gauge theory where the field theory is  conformal and infinitely coupled.  We assume $\bbZ_M$ rotational symmetry for the sake of simplicity, set $\xi=v^M$, $\Phi=\varphi^M$ and $Z=z_0^M$, and consider the polynomials 
\begin{equation}
R=\xi^2-\Phi^2\;,\qquad\qquad S=\xi(\xi-Z)\;.
\end{equation}
We will then study branch curves in the $\xi$ plane. Those in the $v$ plane trivially follow after taking the $M$-th roots of the former ones.
When $\Phi=0$ we can factor out $\xi$ factors in $R$ and $S$, which correspond to regular D3 branes. Then we are left with a short branch curve passing through $Z$ (arising from $R$) and a longer branch curve passing through the origin (arising from $S$) and extending in the nonperturbative region of the dual gauge theory. 
When $\Phi$ does not vanish but is much smaller than the nonperturbative scale in the gauge theory, the longer branch curve is split close to the origin into two branch curves, and in the interior a new branch curve, now arising from $S$, appears passing through the origin. The two long branch curves arising from $R$ extend along the enhan\c con bearing region, whereas the new inner branch curve extends along the new innermost enhan\c con region. What matters for us is that these branch curves are generically disjoint, singling out the H-picture. As shown in Figure \ref{fig:branch_curves_bear_g}, branch curves generically do not meet nor recombine as $\Phi$ varies. Furthermore, it is easy to see that nothing particular happens to the branch curves in the crossover between the perturbative and the nonperturbative region for $\Phi$.  

\begin{figure}[tn]
\centering \includegraphics[width=3cm]{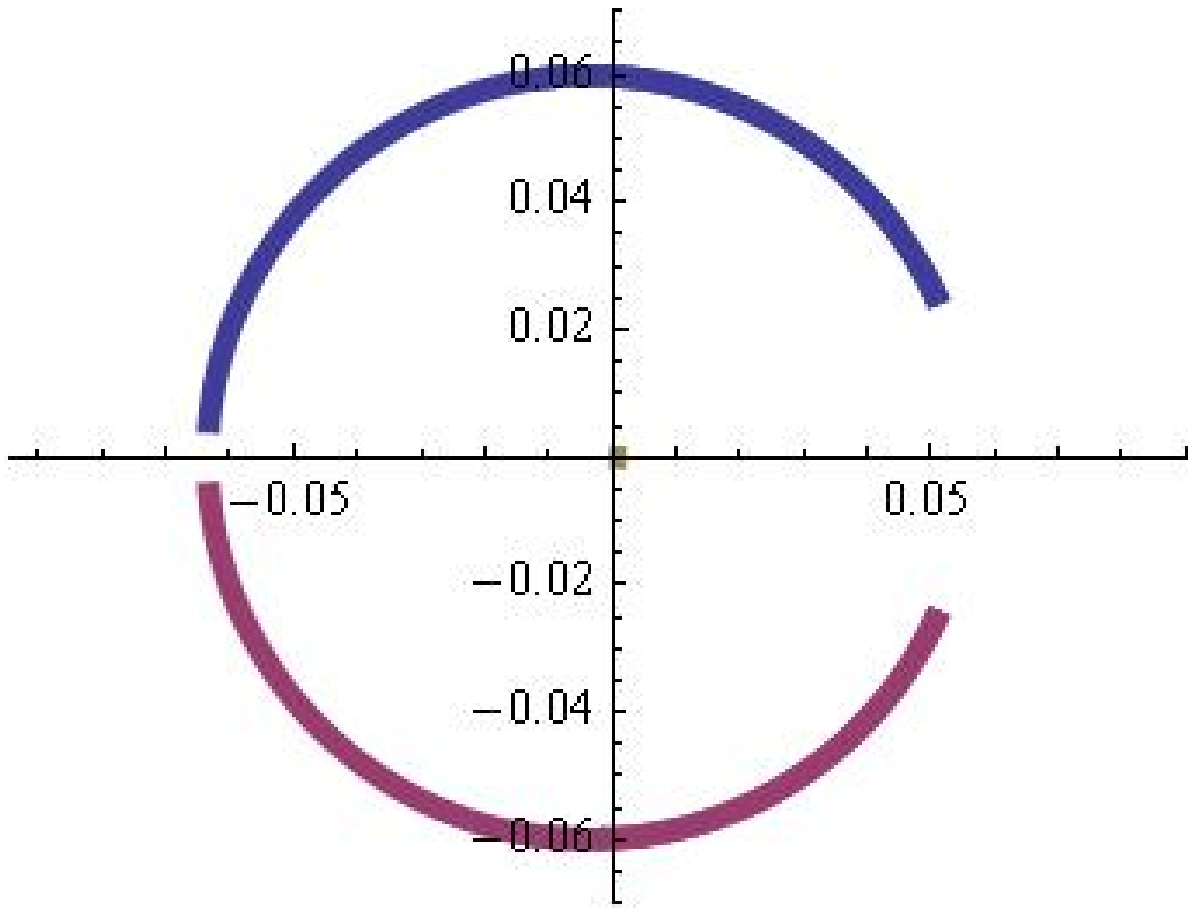}
 \includegraphics[width=3cm]{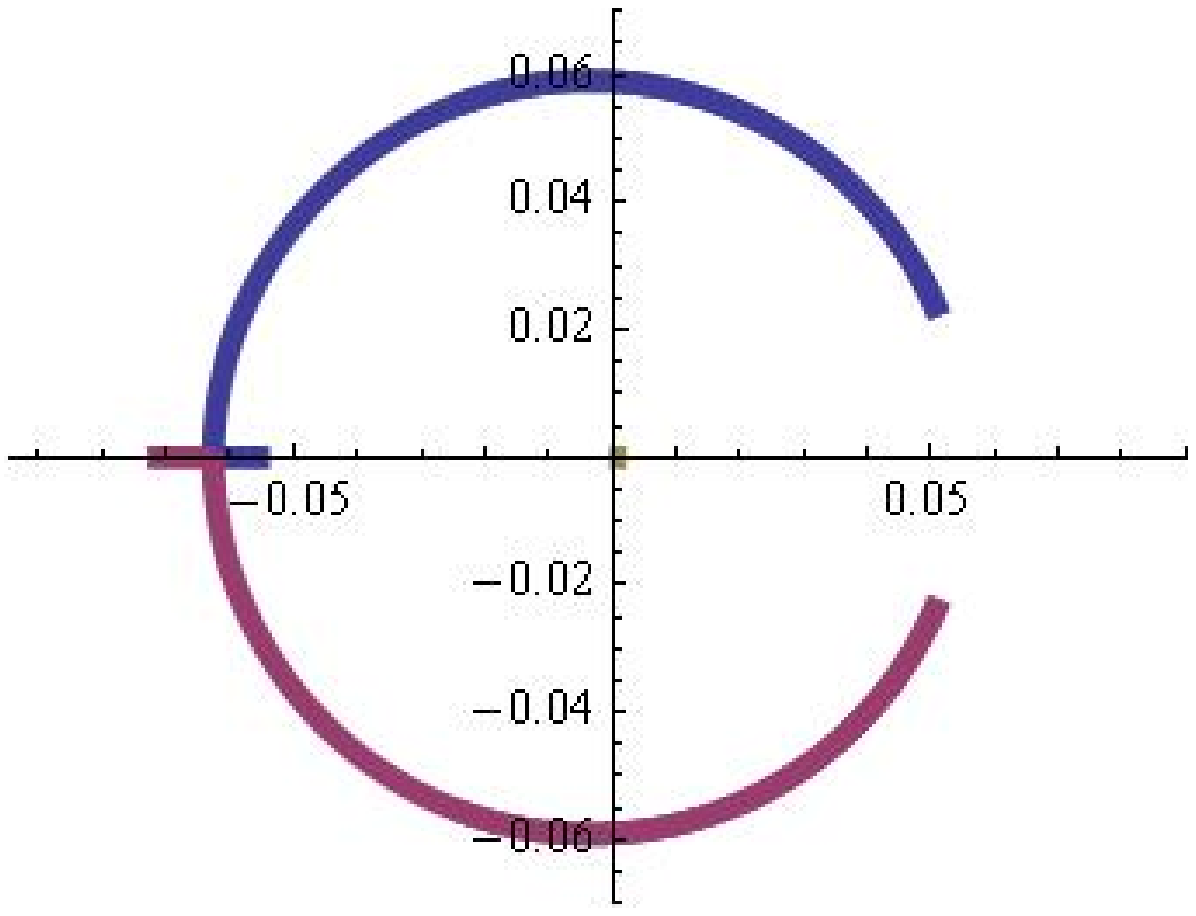}
 \includegraphics[width=3cm]{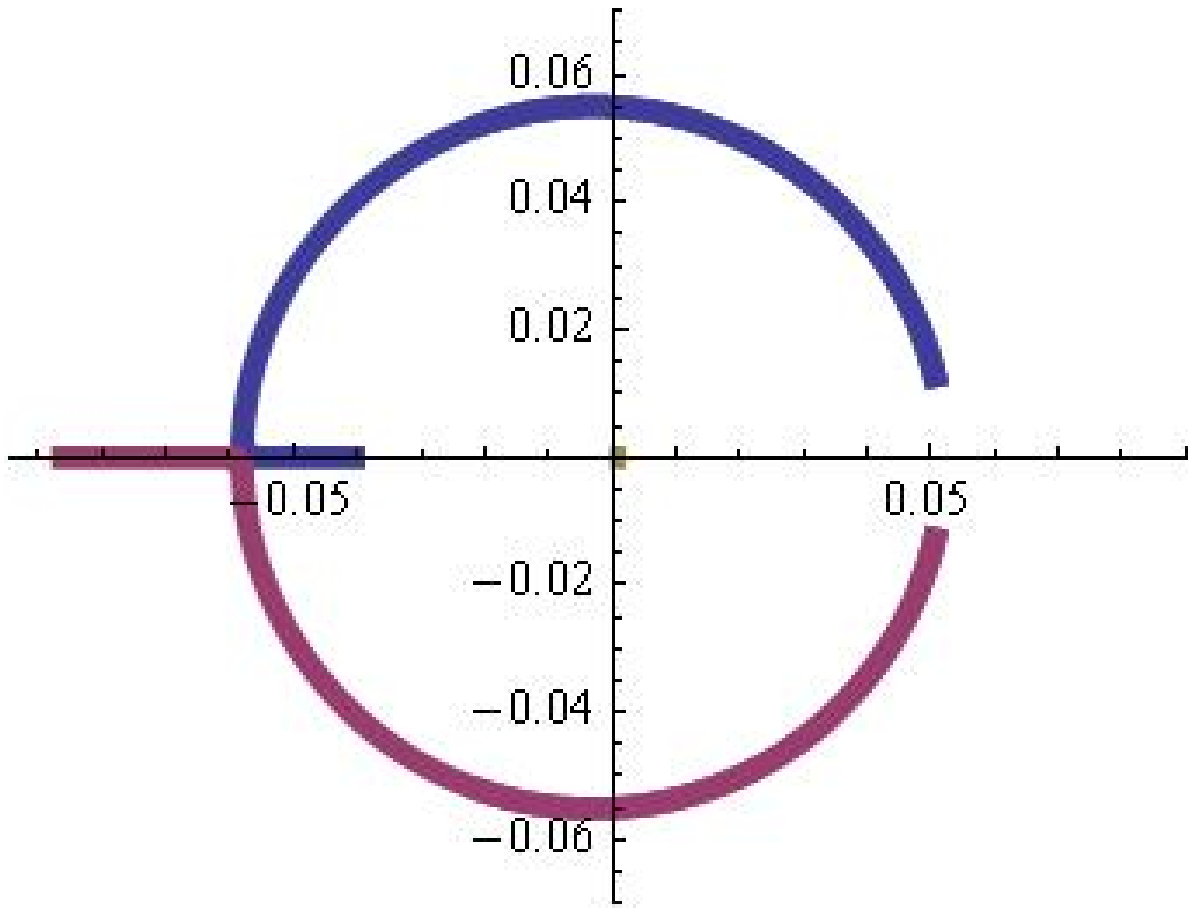}
 \includegraphics[width=3cm]{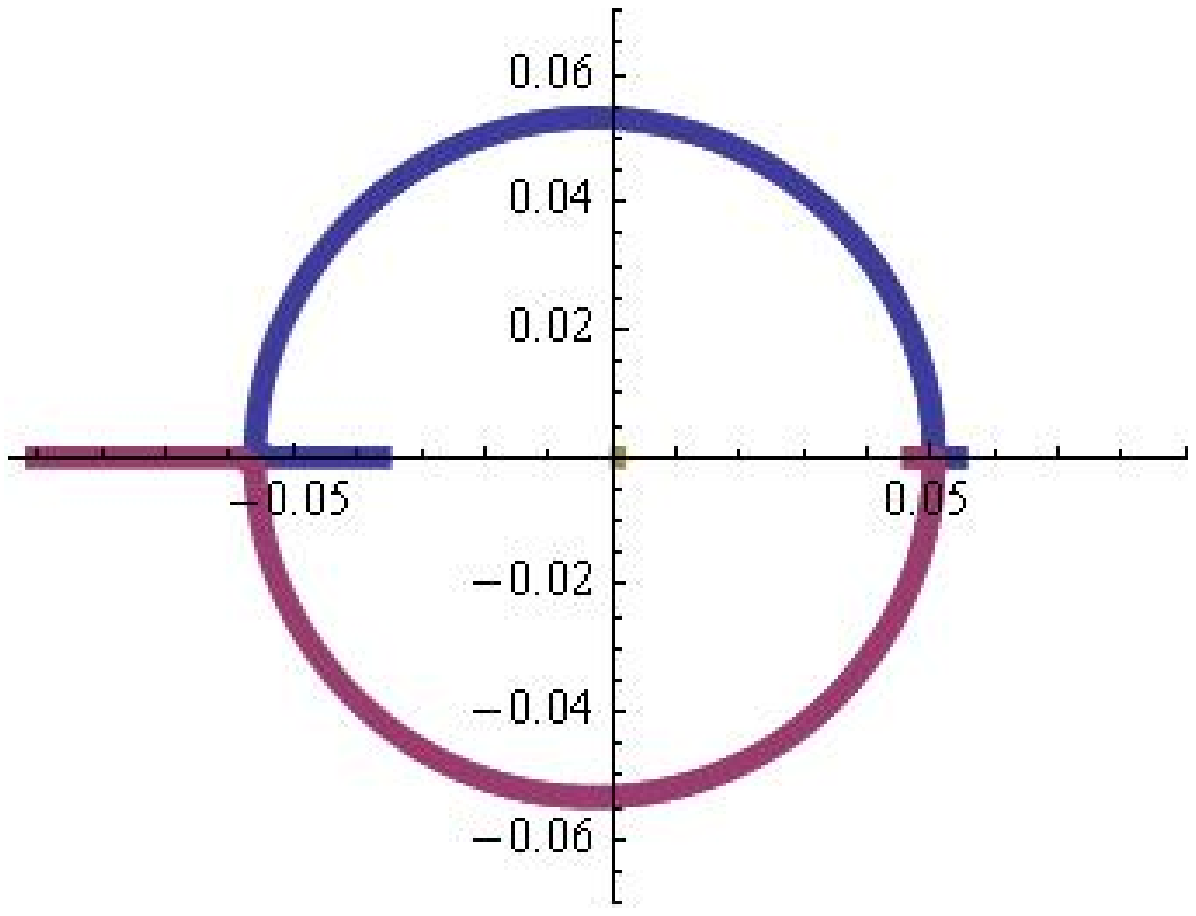}
 \includegraphics[width=3cm]{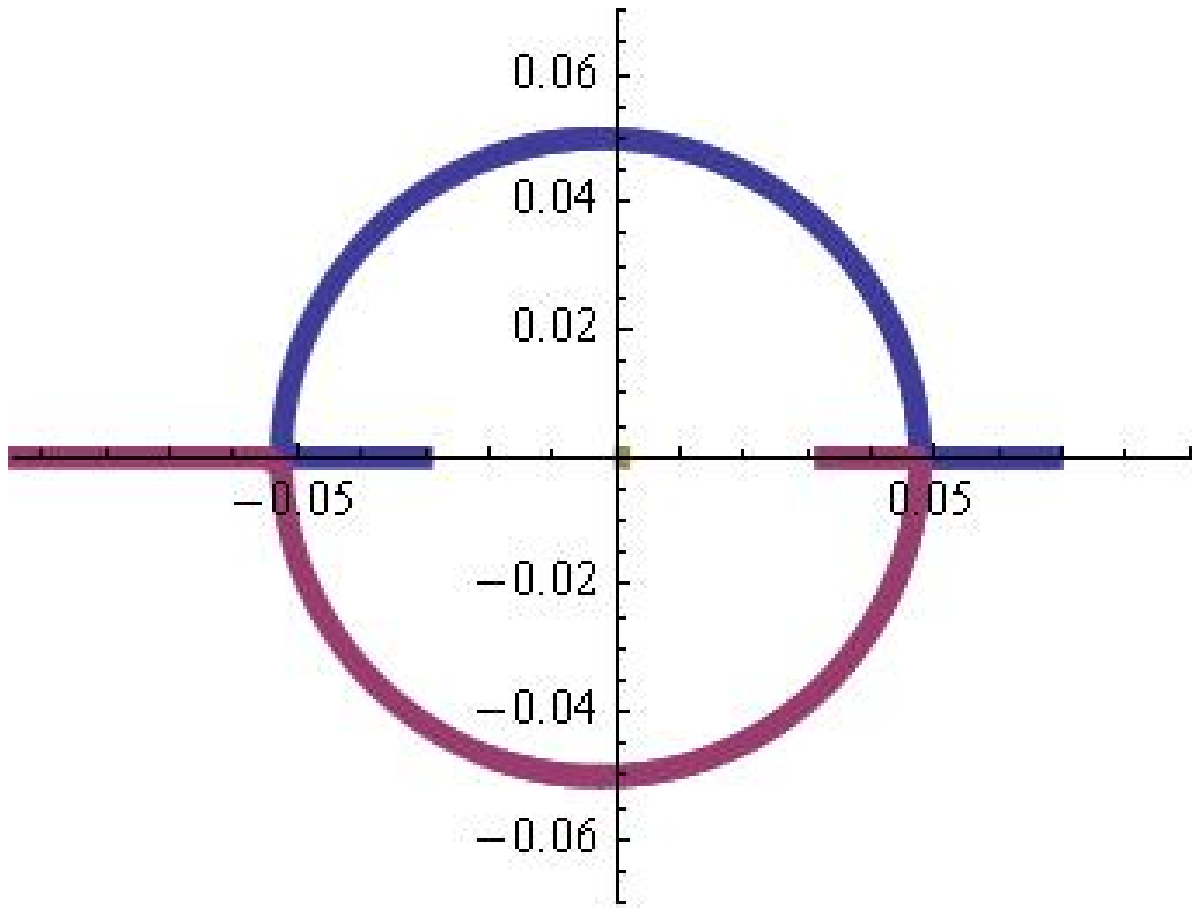}
\caption{\small Branch curves related to the enhan\c con bearing (in magenta and blue) for imaginary $\Phi$, shown in the $\xi$ plane, with $Z,q>0$. $|\Phi|$ decreases from the left to the right, as the radius of the circle shows.  \label{fig:branch_curves_bear}}
\end{figure}
Only at some specific values of $\Phi$ branch points collide: when $\Phi^2=-\frac{Z^2}{4}\frac{g_0(q)^2}{1-g_0(q)}$ the two branch points determined by $R=g_0 S$ coincide, and when $\Phi^2=-\frac{Z^2}{4}\frac{g_0(q)^2}{1+g_0(q)}$ the two branch points determined by $R=-g_0 S$ coincide.

Let us describe what happens, starting with $\Phi$ in the perturbative region, so that in the gauge theory we have a chain of ordinary Higgs breakings $SU(2M)\times SU(2M)\times U(1)\to SU(2M)\times SU(M)\times U(1)^{M+1}\to SU(M)\times U(1)^{3M}$, eventually broken to $U(1)^{4M-1}$ by instantons in the IR. There are a short branch curve passing through $Z$, two short branch curves passing through $\pm \Phi$, and finally a branch curve passing through the origin, associated to the deep IR nonperturbative dynamics. We take $Z$ and $q$ positive for the sake of simplicity, and $\Phi$ imaginary as the coincidence of branch points requires.
As $|\Phi|$ decreases towards a nonperturbative region, the two branch curves passing through $\pm \Phi$ get longer, and close to the nonperturbative regime they form two disjoint arcs on a circle of radius $|\Phi|$. What happens then is depicted in Figure \ref{fig:branch_curves_bear}, where all the branch curves except the one close to $Z$ are shown: the two branch curves with arc shape meet on one side, forming a C shape, which then acquires a bar; then the other sides of the arcs meet, and after that we are left with a circle with two bars on opposite sides. When $|\Phi|$ is further reduced the two bars get longer and the radius of the circle keeps decreasing. Eventually, in the $\Phi\to 0$ limit, the circle disappears, regular D3 branes decouple and we are left with a single long branch curve passing through the origin.
When the two arc-shaped branch curves meet on one side, in the smeared fractional brane approximation one would say that an enhan\c con bearing forms, and then the C-picture could be equivalently used in the interior. However, we see from the picture that there is no domain wall at that point, and the H-picture is still the only valid description of the type IIB solution. 
When the other two sides of the arcs meet too, then we do have a domain wall, but there are also two additional pieces of branch curves originating from the circular part. Then the Dirichlet problems in the interior and in the exterior are independent, but still the H-picture looks preferred. Among other problems that we would face if we insisted in applying the C-picture, the additional pieces of the branch curves would behave like remnants of a noninteger number of fractional branes on the exterior part and a noninteger number of antifractional branes in the interior part, which are joined at a point. That does not seem to be a sensible description. 

There is however an important exception to that statement, that should be clear from the previous discussion. If all the branch points related to a generalized enhan\c con bearing happen to be double, then the branch curves which join pairs of them form a smooth domain wall, without unwanted appendices. Then the C-picture is a sensible description (at least locally). Actually, as we follow a continuous path in the $v$ plane that crosses the branch curve domain wall, the field strength $d\gamma$ picks a minus sign at the wall, so that from this point of view the C-picture  perhaps looks more natural.

Such a coincidence of branch points holds for the vacua that arise from the embedding of the moduli space of $SU(M)$ into that of the infinitely cascading quiver gauge theory. This justifies our use of the C-picture in sections \ref{sec:embedding} and \ref{sec:examples}. Moreover, since we solved the same Dirichlet problems as in the H-picture, up to harmless sign changes and shifts, we are guaranteed that the solution we wrote in the C-picture is correct. Differently stated, one can easily switch to the H-picture by suitably shifting and changing sign to $\gamma$ in \eqref{gamma_generic_poly} at the relevant $b\in \bbZ$ curves.

Finally, following the same rationale of section \eqref{sec:dissolution}, we do not expect any divergence problems in separate contributions to the D3 brane charge and the warp factor when branch points meet. In the case of the infinitely cascading vacua of sections \ref{sec:embedding} and \ref{sec:examples}, that can be readily checked by means of the explicit solution provided in the C-picture.


\section{Summary and conclusions}

In this paper we have shown how to embed the moduli space of $\calN=2$ pure SYM, the gauge theory on a stack of fractional D3 branes at the $A_1$ singularity, into the moduli space of the infinitely cascading quiver gauge theory on regular and fractional D3 branes at the same singularity. Such an embedding, which is provided at the level of Seiberg-Witten curves, along with dualities, allowed us to find an explicit expression for the exact twisted field configuration in type IIB string theory backgrounds dual to this class of vacua, whose SW curves exhibit exactly double branch points, except for at most $2M$ of them which are inherited from SYM. 

The result shows an interesting phenomenon: all fractional D3 branes dissolve into twisted fluxes as soon as the string coupling does not vanish. This is nothing but the dual manifestation of the well known blow-up of suspended D4 branes into M5 branes tubes in type IIA/M theory. An important aspect in the type IIB setting is that this transmutation solves divergence issues in the D3 brane charge and the warp factor, that arise if the naive solutions so far considered in the literature are used.

The same phenomenon holds generally, with and without cascades, for the duals of any points in the moduli space of the quiver gauge theory on any number of D3 branes at the $A_1$ singularity. We have also remarked that in the type IIB solutions, the H-picture for twisted fields, where $b$ is bounded between 0 and 1, is generically singled out as the only valid description of the solution. Only in the case of infinite cascades with exactly double branch points, as those discussed in the first part of the paper, the so called C-picture, where $b$ grows indefinitely towards infinity like in intrinsically $\calN=1$ setups with fractional D3 branes at isolated singularities, is an equally valid description of the same system.

We expect that the phenomena which we studied in this paper appear generally in type IIB duals built out of any systems of D3 branes containing fractional D3 branes of $\calN=2$ kind. We provide an interesting instance of that in the appendix, for an orbifold singular locus with different topology.

\bigskip
\bigskip

\noindent
{\bf \Large Acknowledgements}
\medskip

\noindent We are grateful to Matteo Bertolini and Cyril Closset for comments on the draft.

This research was partly supported by a center of excellence supported by the Israel Science Foundation (grant No. 1468/06), the grant DIP H52 of the German Israel Project Cooperation, the BSF United States-Israel binational science foundation grant 2006157, and the German Israel Foundation (GIF) grant No. 962-94.7/2007.



\appendix



\section{Side remark on cylindrical topology}

We add here a remark on a particular twisted field configuration for fractional D3 branes which are free to move on orbifold fixed loci with cylinder topology rather than plane topology, which is related to the one discussed in subsection \ref{subsec:genus0}. An interesting instance of that is the $\bbZ_2$ orbifold of the deformed conifold studied in \cite{Argurio:2008mt}. 

As an algebraic variety, the $\bbZ_2$ orbifold of the deformed conifold is described by the equation $xy= (z_1 z_2-\eps^2)^2$ in $\bbC^4$. The locus of fixed points under the orbifold action is $x=y=0 \,\wedge \,z_1 z_2=\eps^2$, which indeed has the topology of a cylinder. $\eps$ can be chosen real with no loss of generality.
We can change coordinates by setting 
\begin{equation}
\begin{cases}
z_1= v+u\\
z_2= v-u
\end{cases}\;,
\end{equation}
so that $v^2-u^2=\eps^2$. Therefore we can take
\begin{equation}
\begin{cases}
z_1= v+\sqrt{v^2-\eps^2}\\
z_2= v-\sqrt{v^2-\eps^2}
\end{cases}\;.
\end{equation}
There are two Riemann sheets for $v$: crossing the branch cut of the square root we switch $z_1\leftrightarrow z_2$. 
The circle where the fixed locus intersects the tip of the deformed conifold is $x=y=0$ and $z_1=\eps\,e^{i\alpha}$ (or equivalently $z_2=\eps\,e^{-i\alpha}$), which becomes $v=\eps\cos\alpha$ (or equivalently $u=i\eps\sin\alpha$) in the new coordinates. Such a circle is squeezed into a double segment between $-\eps$ and $\eps$ in the new coordinate $v$. By gluing the two sheets at this double segment we obtain a surface with cylindrical topology.

It is then natural to write a solution for twisted fields analogous to \eqref{gamma_genus_zero} on such a singularity, by taking
\begin{equation}
\gamma= \frac{iP}{\pi}\cosh^{-1} \frac{v}{\eps}= \frac{iP}{\pi}\log\left[\frac{v}{\eps}+\sqrt{\left(\frac{v}{\eps}\right)^2-1}\right]=\frac{iP}{\pi}\log\frac{z_1}{\eps}= 
-\frac{iP}{\pi}\log\frac{z_2}{\eps}\;.
\end{equation}
The previous configuration can be rewritten as 
\begin{equation}
\gamma = \begin{cases}
\frac{iP}{\pi}\log\frac{z_1}{\eps} \qquad \mathrm{if} \qquad |z_1|>|\eps|\\
\frac{iP}{\pi}\log\frac{z_2}{\eps} \qquad \mathrm{if} \qquad |z_2|>|\eps|
\end{cases}\;,
\end{equation}
since we switch from one region to the other of the cylinder as we cross the cut. This is the simplest twisted fields configuration which describes a cascade leading to the $SU(P)\times SU(P) \times SU(P)$ gauge theory studied in \cite{Argurio:2006ny} in the deep IR \cite{Argurio:2008mt}.%
\footnote{These twisted fields have to be supplemented by untwisted 3-form fluxes, as explained in \cite{Argurio:2008mt}.}
In the old language, $P$ tensionless $\calN=2$ fractional D3 branes lie at the circle $|z_1|=|z_2|=\eps$ and account for the charge discontinuity. In the new language, the branch curve which looked like a segment in section \ref{subsec:genus0} becomes here a circle that separates the two branches of the orbifold fixed locus. Curiously, it turns out that the naive logarithm of $z_1$ or $z_2$ here captures the nonperturbative dynamics that make all the branch curves join into loops, including the `infrared' one at the tip of the deformed conifold.


\end{document}